\documentclass[%
 aip,
 jcp,%
 amsmath,amssymb,
 reprint,%
]{revtex4-2}

\usepackage{subfig, graphicx}
\usepackage{verbatim}
\usepackage[version=3]{mhchem}
\usepackage{latexsym}
\usepackage{amsmath}
\usepackage{setspace}
\usepackage[nolist]{acronym}
\usepackage{multirow}
\usepackage{upgreek}
\usepackage{amssymb}
\usepackage{floatrow}
\usepackage{tipa}
\usepackage{dcolumn}
\usepackage{bm}
\usepackage{newunicodechar} 
\usepackage{tabularx}

\newcommand{\SbX}{Sb$_2$X$_3$}
\newcommand{\SbXsp}{Sb$_2$X$_3$ } 
\newcommand{\SbS}{Sb$_2$S$_3$}
\newcommand{\SbSsp}{Sb$_2$S$_3$ }
\newcommand{\SbSe}{Sb$_2$Se$_3$}
\newcommand{\SbSesp}{Sb$_2$Se$_3$ }

\newcommand*{\citen}[1]{%
  \begingroup
    \romannumeral-`\x 
    \setcitestyle{numbers}%
    \cite{#1}%
  \endgroup   
}

\begin{document}

\preprint{AIP/123-QED}

\title{Lone pair driven anisotropy in antimony chalcogenide semiconductors}

\author{Xinwei Wang}
\affiliation{ 
Department of Materials, Imperial College London, Exhibition Road, London SW7 2AZ, UK
}
\author{Zhenzhu Li}
\affiliation{ 
Department of Materials, Imperial College London, Exhibition Road, London SW7 2AZ, UK
}
\affiliation{ 
Department of Materials Science and Engineering, Yonsei University, Seoul 03722, Korea
}
\author{Seán R. Kavanagh}
\affiliation{ 
Department of Materials, Imperial College London, Exhibition Road, London SW7 2AZ, UK
}
\affiliation{ 
Thomas Young Centre and Department of Chemistry, University College London, 20 Gordon Street, London WC1H 0AJ, UK
}
\author{Alex M. Ganose}
\affiliation{ 
Department of Materials, Imperial College London, Exhibition Road, London SW7 2AZ, UK
}
\author{Aron Walsh}
 \email{a.walsh@imperial.ac.uk}
\affiliation{ 
Department of Materials, Imperial College London, Exhibition Road, London SW7 2AZ, UK
}
\affiliation{
Department of Materials Science and Engineering, Yonsei University, Seoul 03722, Korea
}

\date{\today}

\begin{abstract}

Antimony sulfide (\ce{Sb2S3}) and selenide (\ce{Sb2Se3}) have emerged as promising earth-abundant alternatives among thin-film photovoltaic compounds. A distinguishing feature of these materials is their anisotropic crystal structures, which are composed of quasi-one-dimensional (1D) [Sb$_4$X$_6$]$_n$ ribbons. The interaction between ribbons has been reported to be \ac{vdW} in nature and \SbXsp are thus commonly classified in the literature as 1D semiconductors. However, based on first-principles calculations, here we show that inter-ribbon interactions are present in \SbXsp beyond the \ac{vdW} regime. The origin of the anisotropic structures is related to the stereochemical activity of the Sb 5\textit{s} lone pair according to electronic structure analysis. The impacts of structural anisotropy on the electronic, dielectric and optical properties relevant to solar cells are further examined, including the presence of higher dimensional Fermi surfaces for charge carrier transport. Our study provides guidelines for optimising the performance of \SbX-based photovoltaics via device structuring based on the underlying crystal anisotropy.



\end{abstract}

\maketitle

\begin{acronym}
\acro{PV}{photovoltaic}
\acro{CIGS}{copper indium gallium selenide}
\acro{CdTe}{cadmium telluride}
\acro{CZTS}{copper zinc tin sulfide}
\acro{CZTSe}{copper zinc tin selenide}
\acro{CZTSSe}{copper zinc tin sulfoselenide}
\acro{PCE}{power conversion efficiency}
\acro{GBs}{grain boundaries}
\acro{DLTS}{deep-level transient spectroscopy}
\acro{ODLTS}{optical deep-level transient spectroscopy}
\acro{TAS}{thermal admittance spectroscopy}
\acro{ITS}{isothermal transient spectroscopy}
\acro{SQ}{Shockley–Queisser}
\acro{SRH}{Shockley–Read–Hall}
\acro{vdW}{van der Waals}
\acro{VASP}{Vienna Ab initio Simulation Package}
\acro{DFT}{density functional theory}
\acro{LZ}{Lany and Zunger}
\acro{FNV}{Freysoldt, Neugebauer and Van de Walle}
\acro{KO}{Kumagai and Oba}
\acro{1D}{one-dimensional}
\acro{3D}{three-dimensional}
\acro{PDOS}{projected density of states}
\acro{COHP}{crystal orbital Hamilton populations}
\acro{VB}{valence band}
\acro{CB}{conduction band}
\acro{VBM}{valence band maximum}
\acro{CBM}{conduction band maximum}
\acro{PESs}{potential energy surfaces}
\acro{DFPT}{density functional perturbation theory} 
\end{acronym}

\section{Introduction}

Solar \ac{PV} technology which converts the solar energy into electricity provides a clean and sustainable solution to the energy crisis. Current commercial thin-film light absorber materials such as \ac{CdTe} and \ac{CIGS} have achieved certified \ac{PV} efficiencies of over 20\% under laboratory conditions\cite{green2020solar}. However, CdTe suffers from the scarcity of Te elements and toxicity of Cd elements, while in CIGS the cost of In and Ga elements is too high, which limit their large-scale applications. As potential alternatives, antimony chalcogenides (\SbX; X=S, Se) have attracted growing attention. \SbXsp have advantages of long-term stability, optimal bandgaps and high absorption coefficients (\textgreater 10$^{5}$ cm$^{-1}$) with abundant, non-toxic and low-cost constituents\cite{chen2015optical,ghosh1979optical,zhou2014solution}. 
The \ac{PV} conversion efficiencies of \SbXsp devices have increased rapidly during the last decade. The current record efficiencies for pure \SbSsp and \SbSesp have reached 7.5$\%$ and 9.2$\%$ respectively\cite{choi2014highly,li20199}. However, these values are still far below the maximum theoretical efficiency of $\sim$30$\%$ predicted by the \ac{SQ} model\cite{shockley1961detailed} which is an idealised model only based on band gaps of solar absorbers. 

Intensive efforts have been devoted to improve the efficiencies in \SbXsp solar cells. One research direction is the orientation control of \SbXsp films\cite{kim2021importance,hobson2020isotype}.
Based on the understanding that \SbXsp are composed of \ac{1D} [Sb$_4$X$_6$]$_n$ ribbons which are held together by \ac{vdW} forces\cite{caruso2015excitons,song2017highly,guo2018tunable,yang2018adjusting,gusmao2019antimony}, it has been reported that higher \ac{PV} efficiency could be achieved when the \SbXsp films are oriented more perpendicular to the substrate due to more efficient carrier transport along ribbons than between them\cite{zhou2015thin}. 
Consequently, researchers have focused on tailoring the growth orientation of \SbXsp films in order to achieve high efficiencies\cite{yuan2016efficient,wang2017stable,kondrotas2019growth,li20199,zeng2020quasi}. 
However, the origin of anisotropic crystal structures and the connection to physical properties remains unclear. 
Building on recent work that has shown a tolerance to structural reconstructions\cite{mckenna2021self},
understanding how film orientations affect the conversion efficiency in \SbXsp is crucial to designing high-performance devices.

In this study, we investigate the anisotropic structural, electronic and optical properties of \SbXsp using first-principles calculations. The dimensionality of \SbXsp is studied by consideration of the chemical binding energies, carrier effective masses, and Fermi surfaces. Electronic structure analysis further confirms that the anisotropic crystal structures of \SbXsp result from the stereochemical activity of the Sb 5\textit{s} lone pair. Moreover, anisotropic optical properties including dielectric constants and optical absorption spectra are reported.

\section{Methods}
Analysis of the total energy and electronic structure was performed within the framework of Kohn-Sham density-functional theory (DFT)\cite{kohn1965self,dreizler1990density}. Calculations were performed based on DFT as implemented in the \ac{VASP}\cite{kresse1996efficient}. 
The projector augmented-wave (PAW) method\cite{kresse1999ultrasoft} was employed with a plane-wave energy cutoff of 400 eV. All calculations were carried out using the Heyd-Scuseria-Ernzerhof hybrid functional (HSE06)\cite{heyd2003hybrid,krukau2006influence} except for the calculation of ionic contribution of dielectric constants, as a lower-level functional is sufficient for this high-cost calculation\cite{kavanagh2021rapid}.

To account for the weak inter-ribbon interactions, the optB86b-vdW functional\cite{klimevs2011van} was used to calculate the ionic contribution to the static dielectric constant, and the D3 dispersion correction\cite{grimme2004accurate} was used for HSE06 calculations. 
The coefficients used for the zero-damping D3 correction are consistent with the previous research\cite{savory2019complex}. In order to obtain well-converged structures, the atomic positions were optimised until the Hellman-Feynman forces on each atom were below 0.0005 eV \AA$^{-1}$, and the energy convergence criterion was set to 10$^{-8}$ eV. According to convergence tests (shown in Table. S2), the total energies of \SbSsp and \SbSesp converge to within 1 meV/atom at a \textit{k}-point mesh of 7$\times$2$\times$2. Therefore, $\varGamma$-centered \textit{k}-point meshes were set to 7$\times$2$\times$2 for geometry optimisation with primitive unit cells, and 14$\times$4$\times$4 for \ac{PDOS} and \ac{COHP} calculations. For calculations of effective masses, dielectric constants and optical absorption coefficients which are more sensitive to \textit{k}-point density, detailed settings and the proof of convergence are shown in the SI (Table S3-S6).

The crystal structures and partial charge densities were plotted using CrystalMaker$^{\circledR}$\cite{crystalmaker}. The \ac{PDOS} and optical absorption spectra were plotted using the sumo package\cite{ganose2018sumo}. \ac{COHP} calculations were performed using the LOBSTER package\cite{dronskowski1993crystal}. Fermi surfaces were plotted using the IFermi package\cite{ganose2021ifermi}. The conductivity effective mass tensors were calculated by the AMSET package\cite{ganose2021efficient}. The conductivity effective mass (\textit{m}$^{*}$) is obtained based on Boltzmann transport theory\cite{ashcroft1976solid,madsen2006boltztrap} and is defined as\cite{gibbs2017effective}:
\begin{equation}
    \frac{1}{m^*} = \frac{\sigma}{ne^{2}\tau}
\end{equation}
Where $\sigma$ is the conductivity, \textit{n} is the carrier concentration, \textit{e} is the electron charge and $\tau$ is the carrier life time, in this work set to 10$^{-14}$ s. The ionic contribution to the static dielectric constants was calculated by \ac{DFPT}\cite{gajdovs2006linear}, while the electronic part was calculated using the approach developed by Furthm\"uller et al. \cite{gajdovs2006linear} The optical absorption coefficients ($\alpha$) were calculated from the dielectric function by the sumo package\cite{ganose2018sumo} using the equation:
\begin{equation}
    \alpha({\omega}) = \frac{4\pi}{\lambda}\textit{k}({\omega})
\end{equation}
Where $\omega$ and $\lambda$ are the frequency and wavelength of the incident light, respectively. \textit{k}({$\omega$}) is the extinction coefficient defined as
\begin{equation}
    {k}({\omega}) = \sqrt{\frac{\sqrt{\varepsilon_1^2+\varepsilon_2^2}-\varepsilon_1}{2}}
\end{equation}
Where \textit{\textepsilon}$_1$ and \textit{\textepsilon}$_2$ are the real and imaginary parts of the high-frequency dielectric constants, respectively. The figures of distance histograms, COHP, dielectric constants, optical absorption spectra and thickness-dependent maximum efficiencies were plotted using matplotlib\cite{Hunter:2007}.

\section{Results and discussions}
\subsection{Structural properties}
\begin{figure}[h]
    \centering
    {\includegraphics[width=\textwidth]{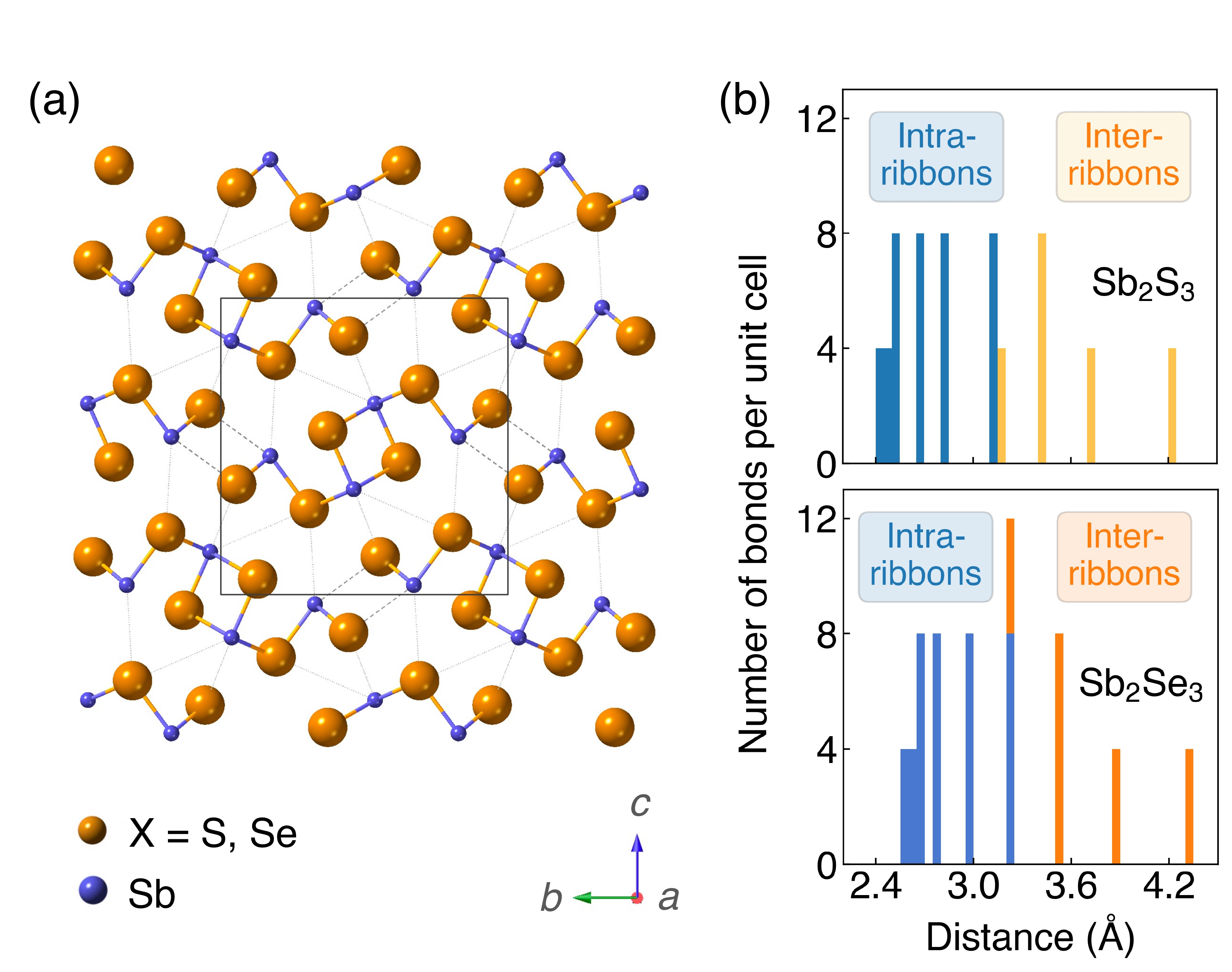}} \\
    \caption{(a) Crystal structures (\textit{Pnma} space group) and (b) histogram of Sb-X distances of \SbSsp and \SbSe. The unit cells are represented by rectangles.}
    \label{fig_structure}
\end{figure}

The ground-state crystal structures of \SbSsp and \SbSesp (\textit{Pnma} space group) are shown in Fig. \ref{fig_structure}a. They are composed of strongly bonded quasi-1D [Sb$_4$X$_6$]$_n$ ribbons units stacked together by weak interactions. In order to better reproduce the crystal structures, different vdW dispersion correction methods were applied in the geometry optimisation process. Lattice parameters calculated by different functionals and different vdW correction methods and obtained by experiments\cite{kyono2002low,savadogo1992studies,salem2001structure,efthimiopoulos2013sb,voutsas1985crystal,hurych1974photoemission} are shown in Table \ref{tab_lattice}. 
The \textit{c} parameter (the direction between ribbons) is significantly overestimated ($>$ 7.0$\%$) with no dispersion correction included for both PBE and HSE06 functionals. Including dispersion corrections, the accuracy of lattice parameter in the \textit{c} direction is greatly improved regardless of the correction methods. 
This demonstrates the presence of significant \ac{vdW} interactions between ribbons, which are much weaker than the bonding within ribbons, and thus the necessity of dispersion corrections when modelling \SbX. For both systems, HSE06+D3 gives the best agreement with experimental measurements (an average difference of 0.7 \% and 0.9 \% for \SbSsp and \SbSesp from the experimental lattice constants, respectively), which is in agreement with previous studies \cite{savory2019complex,stoliaroff2020deciphering}, followed by optB86b performing the next best (1.0 \% and 1.4 \% difference for \SbSsp and \SbSesp, respectively). Thus, the HSE06 functional with the D3 Grimme dispersion correction was used for most of the following calculations and the optB86b functional was used for most convergence tests.

\begin{table*}[ht]
\centering
\resizebox{\textwidth}{!}{%
\begin{tabular}{c@{\extracolsep{0.2cm}}cccccccccccc}
\hline\hline
\multirow{2}{*}{System} & & \multirow{2}{*}{PBE} & \multirow{2}{*}{PBE+D3} & \multirow{2}{*}{PBE+TS} & \multirow{2}{*}{optB86b} & \multirow{2}{*}{HSE06} & \multirow{2}{*}{HSE06+D3} & \multirow{2}{*}{HSE06+TS} & \multicolumn{4}{c}{Experimental data} \\ \cline{10-13} 
\multicolumn{2}{c}{} &  &  &  &  &  &  &  & Ref. [\citen{kyono2002low,savadogo1992studies}] & Ref. [\citen{salem2001structure,efthimiopoulos2013sb}]  & Ref. [\citen{voutsas1985crystal,hurych1974photoemission}]  & Average \\ \hline
\multirow{6}{*}{\ce{Sb2S3}} & \multicolumn{1}{c}{ a } & \begin{tabular}[c]{@{}c@{}}3.87\\ (1.0)\end{tabular} & \begin{tabular}[c]{@{}c@{}}3.84\\ (0.3)\end{tabular} & \begin{tabular}[c]{@{}c@{}}3.88\\ (1.3)\end{tabular} & \begin{tabular}[c]{@{}c@{}}3.86\\ (0.8)\end{tabular} & \begin{tabular}[c]{@{}c@{}}3.80 \\ (-0.8)\end{tabular} & \begin{tabular}[c]{@{}c@{}}3.80 \\ (-0.8)\end{tabular} & \begin{tabular}[c]{@{}c@{}}3.81 \\ (-0.5)\end{tabular} & 3.84 & 3.82 & 3.84 & 3.83 \\ [2ex] 
 & \multicolumn{1}{c}{ b } & \begin{tabular}[c]{@{}c@{}}11.22\\ (-0.4)\end{tabular} & \begin{tabular}[c]{@{}c@{}}10.92\\ (-3.1)\end{tabular} & \begin{tabular}[c]{@{}c@{}}11.09\\ (-1.5)\end{tabular} & \begin{tabular}[c]{@{}c@{}}11.04\\ (-2.0)\end{tabular} & \begin{tabular}[c]{@{}c@{}}11.39 \\ (1.1)\end{tabular} & \begin{tabular}[c]{@{}c@{}}11.20 \\ (-0.5)\end{tabular} & \begin{tabular}[c]{@{}c@{}}11.22 \\ (-0.4)\end{tabular} & 11.22 & 11.27 & 11.29 & 11.26 \\ [2ex] 
 & \multicolumn{1}{c}{ c } & \begin{tabular}[c]{@{}c@{}}12.14\\ (7.0)\end{tabular} & \begin{tabular}[c]{@{}c@{}}11.15\\ (-1.3)\end{tabular} & \begin{tabular}[c]{@{}c@{}}11.54\\ (2.2)\end{tabular} & \begin{tabular}[c]{@{}c@{}}11.32\\ (0.3)\end{tabular} & \begin{tabular}[c]{@{}c@{}}12.09 \\ (6.6)\end{tabular} & \begin{tabular}[c]{@{}c@{}}11.39 \\ (0.9)\end{tabular} & \begin{tabular}[c]{@{}c@{}}11.54 \\ (2.2)\end{tabular} & 11.31 & 11.30 & 11.27 & 11.29 \\ [3ex]
\multirow{6}{*}{\ce{Sb2Se3}} & \multicolumn{1}{c}{ a } & \begin{tabular}[c]{@{}c@{}}4.03\\ (1.2)\end{tabular} & \begin{tabular}[c]{@{}c@{}}3.99\\ (0.3)\end{tabular} & \begin{tabular}[c]{@{}c@{}}4.04\\ (1.5)\end{tabular} & \begin{tabular}[c]{@{}c@{}}4.02\\ (1.0)\end{tabular} & \begin{tabular}[c]{@{}c@{}}3.96 \\ (-0.5)\end{tabular} & \begin{tabular}[c]{@{}c@{}}3.95 \\ (-0.8)\end{tabular} & \begin{tabular}[c]{@{}c@{}}3.97 \\ (-0.3)\end{tabular} & 3.98 & 3.99 & 3.96 & 3.98 \\ [2ex] 
 & \multicolumn{1}{c}{ b } & \begin{tabular}[c]{@{}c@{}}11.53\\ (-1.0)\end{tabular} & \begin{tabular}[c]{@{}c@{}}11.33\\ (-2.7)\end{tabular} & \begin{tabular}[c]{@{}c@{}}11.41\\ (-2.0)\end{tabular} & \begin{tabular}[c]{@{}c@{}}11.46\\ (-1.6)\end{tabular} & \begin{tabular}[c]{@{}c@{}}11.73 \\ (0.8)\end{tabular} & \begin{tabular}[c]{@{}c@{}}11.55 \\ (-0.8)\end{tabular} & \begin{tabular}[c]{@{}c@{}}11.54 \\ (-0.9)\end{tabular} & 11.65 & 11.65 & 11.62 & 11.64 \\ [2ex] 
 & \multicolumn{1}{c}{ c } & \begin{tabular}[c]{@{}c@{}}12.84\\ (8.2)\end{tabular} & \begin{tabular}[c]{@{}c@{}}11.68\\ (-0.9)\end{tabular} & \begin{tabular}[c]{@{}c@{}}12.31\\ (4.2)\end{tabular} & \begin{tabular}[c]{@{}c@{}}11.90\\ (0.9)\end{tabular} & \begin{tabular}[c]{@{}c@{}}12.65 \\ (6.8)\end{tabular} & \begin{tabular}[c]{@{}c@{}}11.93 \\ (1.2)\end{tabular} & \begin{tabular}[c]{@{}c@{}}12.18 \\ (3.2)\end{tabular} & 11.80 & 11.79 & 11.77 & 11.79 \\ \hline\hline

\end{tabular}%
}
\caption{Lattice parameters (\AA) of \SbSsp and \SbSesp as calculated by different functionals and different vdW dispersion correction methods. The percentage error (\%) relative to the experimental average is given in parentheses}
\label{tab_lattice}
\end{table*}

Histograms of distances between Sb and X ions are shown in Fig. \ref{fig_structure}b. The widely distributed bond lengths indicate the anisotropic connectivity of \SbX. The bond lengths of \SbSesp are slightly larger than those of \SbSsp due to the increased anion radius. 
In the study of Deringer et al\cite{deringer2015vibrational}, analysis of the integrated COHP and the harmonic force constants showed a clear separation between strong intra-ribbon and weaker inter-ribbon interactions in \ce{Sb2Se3}.
In order to quantify the interaction between ribbons, inter-ribbon binding energies ($E_\textrm{b}$) (per unit cell) along the \textit{b} and \textit{c} directions are calculated by
\begin{equation}
    E_{\textrm{b}(\textrm{b})} = (E_\textrm{1D}-E_\textrm{2D})/2
\end{equation}  
\begin{equation}
    E_{\textrm{b}(\textrm{c})} = (2E_\textrm{2D}-E_\textrm{t})/N
\label{eq_Ebc}
\end{equation}
where $E_\textrm{1D}$ and $E_\textrm{2D}$ are total energies of one 1D [Sb$_4$X$_6$]$_n$ ribbon in isolation and one 2D [Sb$_4$X$_6$]$_n$ ribbon periodically repeated along \textit{a} and \textit{b} directions, respectively (the 1D and 2D substructures are given in Fig. S1). $E_\textrm{t}$ is the total energy of the unit cell. The parameter 2 in Eq. \ref{eq_Ebc} is due to the fact that one unit cell of \SbXsp contains two [Sb$_4$X$_6$]$_n$ ribbons. Note that the substructures were directly taken from the optimised structures and kept unrelaxed in order to avoid structural distortion effects, as is typical for binding energy calculations\cite{mounet2018two,bjorkman2012van}. The effect of optimisation was also tested and the $E_\textrm{b}$ was $\sim$0.1 eV lower after the relaxation of substructures, which does not qualitatively influence the results. The choice of \textit{N} depends on whether $E_{\textrm{b}(\textrm{c})}$ is defined as per atom or per bond. Table \ref{tab_binding} shows the calculated $E_\textrm{b}$ using the HSE06 functional and D3 dispersion correction. It can be seen that $E_\textrm{b}$ of \SbSsp and \SbSesp are both over 10 kJ·mol$^{-1}$, which are both beyond the typical vdW regime (0.4 $\sim$ 4 kJ·mol$^{-1}$)\cite{garrett1999biochemistry}. This is consistent with previous research that the distance of Sb-S between ribbons in \SbSsp is shorter than the sum of Sb and S vdW radii at 293 K\cite{kyono2002low}. Moreover, $E_{\textrm{b}(\textrm{b})}$ is larger than $E_{\textrm{b}(\textrm{c})}$ due to the elongation of ribbons along \textit{b}, and the $E_\textrm{b}$ of \SbSesp is slightly larger than that of \SbS. The calculated values agree well with previous calculations\cite{filip2013g} and indicate that the inter-ribbon interactions of \SbSsp and \SbSesp are both between the vdW and covalent regime. The results of $E_\textrm{b}$ without vdW corrections are given in Table S1.

\begin{table}[h]
\begin{ruledtabular}
\begin{tabular}{cccc}
\multirow{2}{*}{System} & $E_{\textrm{b}(\textrm{b})}$        & \multicolumn{2}{c}{$E_{\textrm{b}(\textrm{c})}$}            \\ \cline{2-4} 
                  & per bond & per bond (N=16) & per atom (N=20) \\ \hline
Sb$_2$S$_3$                 & 27.44    & 15.96           & 12.77           \\ 
Sb$_2$Se$_3$                & 31.13    & 17.95           & 14.36           \\ 
\end{tabular}
\end{ruledtabular}
\caption{Inter-ribbon binding energies (kJ·mol$^{-1}$) of \SbSsp and \SbSe}
\label{tab_binding}
\end{table}

\subsection{Electronic properties}

The strength of the interaction between ribbons is closely related to the distortion of Sb which originates from the stereochemically active Sb 5\textit{s} lone pairs. Before going further into the lone pair analysis, the density of states and orbital overlaps are first investigated.

\begin{figure}[h]
    \centering
    {\includegraphics[width=\textwidth]{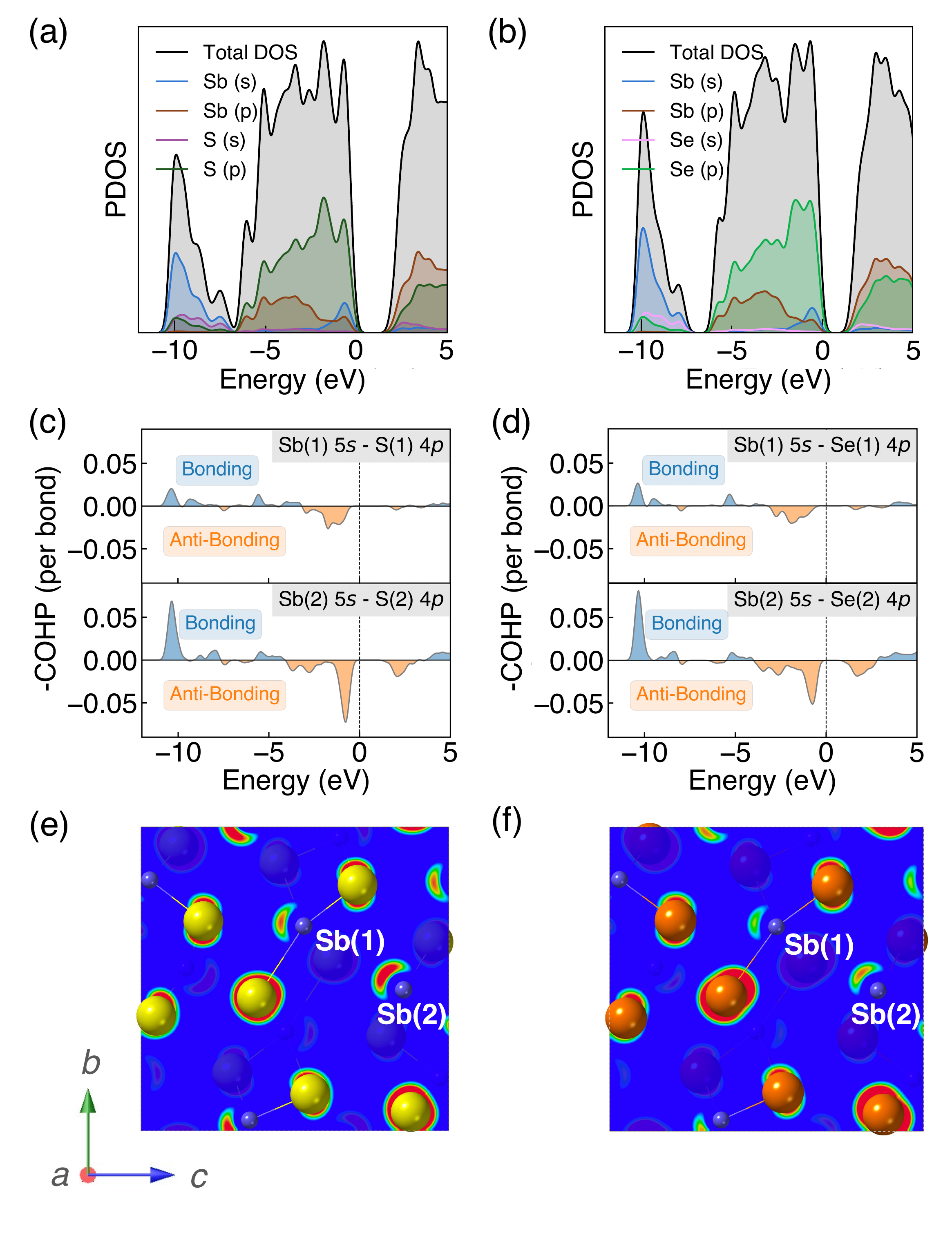}} 
    \caption{(a)-(b) Projected density of states (PDOS), (c)-(d) crystal orbital Hamilton populations (COHP) and (e)-(f) partial charge densities of antimony chacogenides. The range of isosurface value for partial charge densities is set to 0.05 $\sim$ 0.10 \textit{e}/\AA$^3$ for both \SbSsp and \SbSe. The \ac{VBM} is set to 0.} 
    \label{fig_electronic}
\end{figure}

The PDOS are shown in Fig. \ref{fig_electronic}a and \ref{fig_electronic}b. The \ac{VB} can be divided into two main parts. The highest occupied VB between -1 and 0 eV consists of S 3\textit{p}/Se 4\textit{p}, Sb 5\textit{s} and Sb 5\textit{p} states. While the states between -10 and -7 eV mainly consists of Sb 5\textit{s} orbitals, alongside some small contributions from S 3\textit{s}/Se 4\textit{s} and S 3\textit{p}/Se 4\textit{p} states. A valley at about 2 eV below the valence band maximum (VBM) is demonstrated to be one of the major characteristics of energy distribution curves for \SbSesp according to the photoemission measurements\cite{hurych1974photoemission}. Our calculated PDOS of \SbSesp also shows a valley at $\sim$ -2 eV which is in good agreement with the experimental results. The \ac{CB} are dominated by Sb 5\textit{p} and S 3\textit{p}/Se 4\textit{p} states. These results agree well with earlier studies of PDOS on \SbX \cite{caracas2005first,kocc2012first,deringer2015vibrational,radzwan2017first}.

The bonding and antibonding interactions are further studied by COHP\cite{dronskowski1993crystal} (shown in Fig. \ref{fig_electronic}c and \ref{fig_electronic}d). Two separate cases are plotted since Sb has two distinct chemical environment. The interaction is weaker in Sb(1)-X(1) than Sb(2)-X(2) which agrees with the longer bond lengths of Sb(1)-X(1). Combined with the results of PDOS, the energy range from -10 to -7 eV corresponds to a bonding interaction between Sb 5\textit{s} and S 3\textit{p}/Se 4\textit{p} states, whereas the region at the top of the VB corresponds to an antibonding state, which is similar to other quasi-1D systems with stereochemically active lone pairs \cite{ganose2016relativistic}. The photoemission measurements for \SbSesp \cite{hurych1974photoemission} show that the lower part of VB below $\sim$ -6 eV is contributed by bonding states which agrees well with our results. Moreover, the interaction is stronger in \SbSsp than \SbSesp which will be discussed in detail later. 

\begin{figure*}
    \centering
    \includegraphics[width=\textwidth]{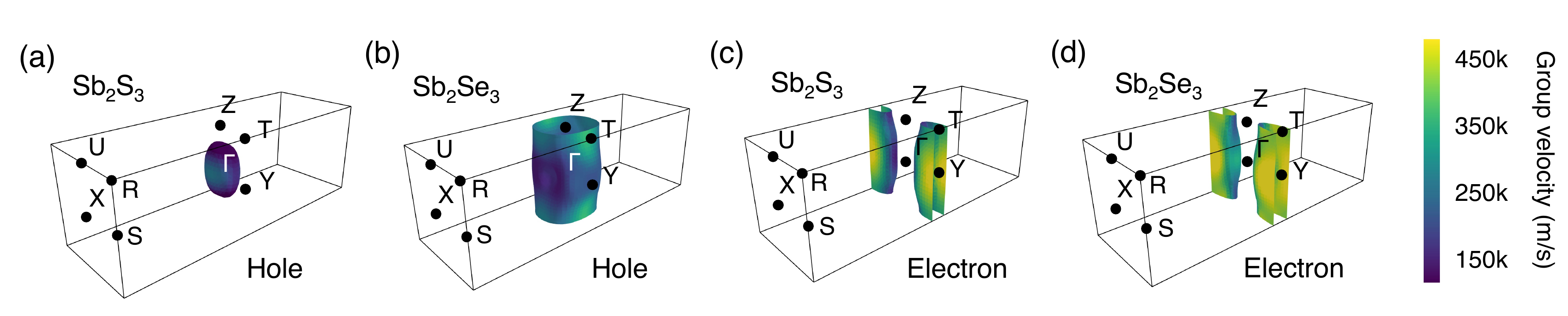}
    \caption{Fermi surfaces of \SbSsp and \SbSe. (a) and (c) are hole Fermi surfaces (0.1 eV below the valence band maximum (VBM)), while (b) and (d) are electron Fermi surfaces (0.1 eV above the conduction band minimum (CBM)). The different colors represent the magnitude of group velocity (m/s)}
   \label{fig_fermi}
\end{figure*}

The different inter-ribbon interactions in \SbSsp and \SbSesp result from the Sb 5\textit{s} lone pair formation. In the revised lone pair model\cite{walsh2011stereochemistry}, the interaction between the antibonding states of cation \textit{s} - anion \textit{p} and cation \textit{p} states results in the formation of stereochemically active lone pairs. Smaller energy difference between cation \textit{s} and anion \textit{p} states will facilitate stronger interaction and thus stronger lone pair formation. In our systems, for the Sb(III) oxidation state found in \ce{Sb2X3}, the formal electronic configuration of Sb is 5\textit{s}$^2$5\textit{p}$^0$. Based on the discussions above, the Sb 5\textit{s} states interact with the S 3\textit{p}/Se 4\textit{p} states in the VB forming filled bonding and anti-bonding states.
The additional interaction of the nominally empty Sb 5\textit{p} orbitals stabilises the system by lowering the total energy, which is similar to other lone pair systems\cite{walsh2005origin,walsh2011stereochemistry,allen2013electronic,kavanagh2021hidden}. 
The resulting stereochemically active lone pair results in an asymmetric electronic density at the top of VB which can be visualised by the contour plot of partial charge density. Partial charge densities for the states between -1 and 0 eV (with respect to the \ac{VBM}) are shown in Fig. \ref{fig_electronic}e and \ref{fig_electronic}f. They are obtained by cutting the (100) plane through Sb atoms. The lone pair is stronger in \SbSsp compared with \SbSesp due to the smaller energy separation and increased overlap of Sb 5\textit{s} and S 3\textit{p}. 

To connect the electronic structure to transport properties, effective masses of electrons and holes were calculated (shown in Table \ref{tab_effective}). According to the electronic band structures of \SbSsp and \SbSesp (shown in Fig. S2), the band dispersions around the extrema are relatively flat which are far from parabolic, and several extrema in different parts of the Brillouin Zone which are close in energy could be involved in contributing to conductivity. Therefore, effective masses in \SbXsp are quite sensitive to calculation parameters (particularly the \textit{k}-point density), and effective masses larger than 2 are rounded to the nearest whole numbers in Table \ref{tab_effective} and S3. The harmonic mean is used to average the values for a polycrystalline sample with random orientations, and the anisotropy ratio ($\textit{a}_r$) is defined as the ratio of maximum to minimum effective mass.  The average effective masses of holes are larger than those of electrons for both \SbSsp and \SbSe, indicating that \SbSsp and \SbSesp may be better n-type semiconductors. Nevertheless, the electron effective masses show a stronger anisotropy. $\textit{a}_r$ is larger in \SbSesp than \SbSsp for both electron and hole effective masses, suggesting \SbSesp has stronger anisotropy which agrees with the weaker lone pair in \SbSe. The electron and hole effective masses are the largest along [001] for both \SbSsp and \SbSe, indicating the conductivity between ribbons will be lower than along the other two directions. In general the values follow the trend \textit{x} $<$ \textit{y} $<$ \textit{z}, with the exception of the hole effective mass for \ce{Sb2Se3}, which is discussed later. Moreover, the 2D nature of transport (with small effective masses in two directions and large effective mass in the other direction) is similar to other so-called “quasi-1D” systems such as BiSI and BiSeI\cite{ganose2018defect}. Our calculated effective masses deviate largely from other studies on \SbX\cite{nasr2016first,qiu2019crystal}. One possible reason of the discrepancy could be the use of simple parabolic fitting or the consideration of solely the $\varGamma$ point in other computational investigations. The choice of functionals could be another important factor, as demonstrated by Whalley et al. \cite{whalley2019impact}  It has been demonstrated that semi-local functionals would not only underestimate the bandgap, but also would influence the shape of band structures, resulting in overestimated nonparabolicity.

\begin{table}[h]
\begin{ruledtabular}
\begin{tabular}{ccccccc}

{System} &{} & \textit{x}    & \textit{y}    & \textit{z}    & $\overline{m}^{*}$ &$\textit{a}_r$ \\ \hline
\multirow{2}{*}{Sb$_2$S$_3$}        & \textit{m}$_{e}^{*}$/\textit{m}$_{0}$       & 0.16     & 0.92     & 5     & 0.40  &31.25 \\ 
                          & {m}$_{h}^{*}$/\textit{m}$_{0}$       & 0.47     & 0.65    & 0.97     & 0.64 &2.06 \\ [1ex]
\multirow{2}{*}{Sb$_2$Se$_3$}       & {m}$_{e}^{*}$/\textit{m}$_{0}$       & 0.14  & 0.81  & 7  & 0.35 &50.00\\ 
                          & {m}$_{h}^{*}$/\textit{m}$_{0}$       & 0.85 & 0.55 & 3 & 0.90 &5.45\\ 
\end{tabular}
\end{ruledtabular}
\caption{Effective masses of \SbSsp and \SbSe. The harmonic mean is represented by $\overline{m}^{*}$. The anisotropy ratio ($\textit{a}_r$) is defined as the ratio of maximum to minimum effective mass.}
\label{tab_effective}
\end{table}

To further illustrate the dimensionality of the electronic structure, Fermi surfaces were plotted at 0.1 eV below (above) the VBM (CBM) using the IFermi package\cite{ganose2021ifermi} (shown in Fig. \ref{fig_fermi}). 0.1 eV is an arbitrary value intended to indicate the shape of the Fermi surface close to the band edge. Due to the tails of the Fermi-Dirac distribution, this energy range will be occupied at room temperature and the Fermi surface is therefore reflective of the states that govern transport properties. The Fermi surfaces of 0.08 and 0.12 eV below (above) the valence band maximum (conduction band maximum) were also shown in Fig. S3 and S4 which qualitatively show the same behaviour. An ellipsoidal Fermi surface is found for holes in \ce{Sb2S3} (Fig. \ref{fig_fermi}a), indicating dispersion in three dimensions\cite{albert2015torque}. 
The shape of elliptical cylinders found for electrons in \ce{Sb2S3} (Fig. \ref{fig_fermi}c) can be classified as quasi-2D with small dispersion in the [001] direction\cite{albert2015torque}. 
These agree well with observation that the hole effective mass of \SbSsp is much smaller than the electron effective mass in the [001] direction. 
In contrast, the hole and electron Fermi surfaces of \SbSesp are both quasi-2D (shown in Fig. \ref{fig_fermi}b and \ref{fig_fermi}d). 
It can be seen that the electron Fermi surfaces of \SbSsp and \SbSesp have similar topology, which indicates similar transport behaviour, whereas their hole Fermi surfaces have a significant difference in terms of the dimensionality. 
Indeed, for \ce{Sb2S3} the three components of the effective mass are all below one, while for \ce{Sb2Se3} the z component is greater than 3.
We link this behaviour to the stronger lone pair distortion of \SbSsp and the resulting shorter inter-ribbon Sb-S bonds along the [001] direction.

\subsection{Optical properties}

The dielectric constants are important descriptors for the optical properties of crystals. 
The static dielectric constant (\textit{\textepsilon}$_0$) is defined as the sum of the ionic and high-frequency response to an external electric field. The complex dielectric function \textit{\textepsilon}$(\omega)$ is given by:
\begin{equation}
\varepsilon(\omega) = \varepsilon_{1}(\omega) + i\varepsilon_{2}(\omega)
\end{equation}
Where \textit{\textepsilon}$_1$ and \textit{\textepsilon}$_2$ are the real part and imaginary part of the frequency-dependent dielectric function, respectively. For orthorhombic structures, the dielectric tensor has three distinct non-zero components. 
As shown in Table \ref{tab_dielectric}, the dielectric constants of \SbXsp are anisotropic and relatively large, which is common in lone-pair containing crystals\cite{kavanagh2021hidden,ganose2016relativistic}. Large dielectric constants indicate the potential for strong screening to charged defects and low recombination losses\cite{zeng2016antimony,walsh2017instilling}.  It can be seen that the dielectric constants are larger in the \textit{x} and \textit{y} directions than the \textit{z} direction, indicating the screening is stronger in the \textit{ab} plane. Moreover, the dielectric screening in \SbXsp is shown to be dominated by the lattice polarization as the ionic contribution is much larger than the electronic contribution. The large ionic dielectric constants can be attributed to large Born effective charges in \SbX \cite{liu2014first,cheng2019understanding}.
The anisotropy ratio ($\textit{a}_r$) (defined as the ratio of maximum to minimum dielectric constant) is larger in \SbSesp than \SbSsp for both static and high-frequency dielectric constants, indicating \SbSesp has stronger anisotropy which is consistent with previous discussions.

\begin{figure}[ht]
    \centering
    \includegraphics[width=\textwidth]{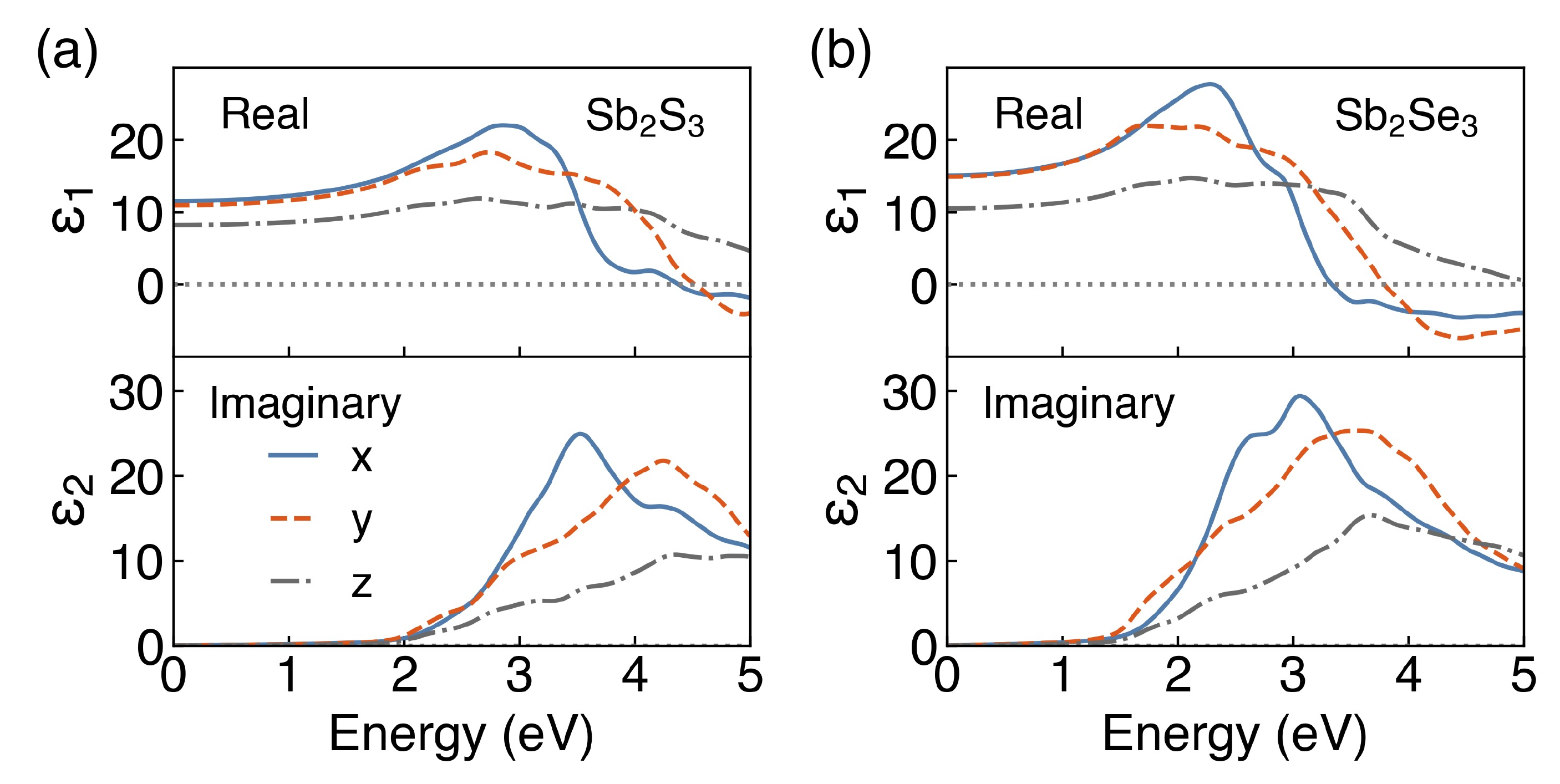}
    \caption{Calculated high-frequency dielectric functions of (a) \SbSsp and (b) \SbSe.}
    \label{fig_dielectric}
\end{figure}

The real and imaginary parts of the high-frequency (\textit{\textepsilon}$_{\infty}$) dielectric functions are plotted in Fig. \ref{fig_dielectric}. Combined with the results of PDOS, the peaks in the imaginary parts of dielectric functions mainly correspond to the optical transition from the S 3\textit{p}/Se 4\textit{p} states in the valence band to the Sb 5\textit{p} states in the conduction band. Our calculated dielectric constants are in excellent agreement with ellipsometry measurements on polycrystalline thin films ($\epsilon_{\infty, x}, \epsilon_{\infty, y}$ and $\epsilon_{\infty, z}$ of 12.5, 10.8 and 7.0 for \SbSsp\cite{schubert2004generalized}, respectively, and an averaged $\epsilon_{\infty}$ of 14.3 for \SbSesp \cite{chen2015optical}). Furthermore, our results are inline with previous theoretical studies \cite{nasr2011electronic,lakhdar2014dielectric,kocc2012first,maghraoui2013synthesis,lawal2018investigation}.

\begin{table}[h]
\begin{ruledtabular}
\begin{tabular}{ccccccccc}
\multirow{2}{*}{System} & \multicolumn{4}{c}{\textit{\textepsilon}$_0$} & \multicolumn{4}{c}{\textit{\textepsilon}$_{\infty}$} \\ \cline{2-9} 
                                    & \textit{x}     & \textit{y}      & \textit{z}   &$\textit{a}_r$  & \textit{x}      & \textit{y}     & \textit{z}   &$\textit{a}_r$  \\ \hline
Sb$_2$S$_3$                               & 98.94 & 94.21  & 13.14 &7.53 & 11.55      & 10.97     & 8.25  &1.40   \\ 
Sb$_2$Se$_3$                              & 85.64 & 128.18 & 15.00 &8.54 & 15.11  & 14.92 & 10.53 &1.43\\ 
\end{tabular}
\end{ruledtabular}
\caption{Calculated static (\textit{\textepsilon}$_0$) and high-frequency (\textit{\textepsilon}$_{\infty}$) dielectric constants of \SbSsp and \SbSe. The anisotropy ratio ($\textit{a}_r$) is defined as the ratio of maximum to minimum dielectric constant.}
\label{tab_dielectric}
\end{table}

\begin{figure}[ht]
    \centering
    \includegraphics[width=\textwidth]{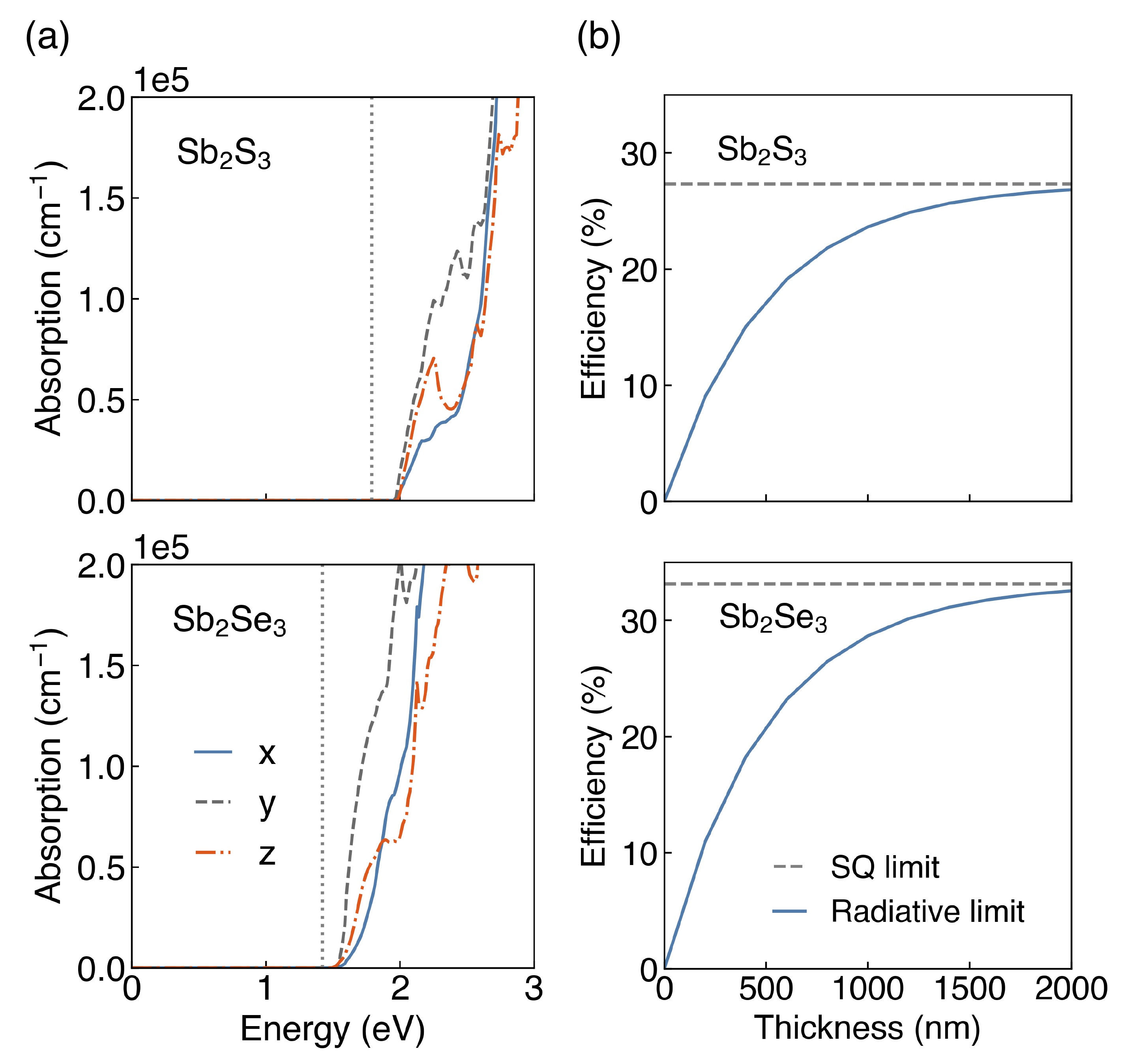} 
    \caption{(a) Calculated optical absorption spectra of \SbSsp and  \SbSe. The fundamental band gaps are shown in grey dotted lines. (b) Thickness-dependent maximum efficiencies based on the radiative limit of \SbSsp and \SbSe.}
    
    \label{fig_optical}
\end{figure}

The calculated optical absorption spectra, and upper limit to photovoltaic efficiency, for \SbSsp and \SbSesp are shown in Fig. \ref{fig_optical}. As can be seen in Fig. \ref{fig_optical}a, both \SbSsp and \SbSesp possess high optical absorption coefficients above the band edge (in the range of 10$^{5}$ cm$^{-1}$) which could more effectively absorb photons and generate electron-hole pairs. This agrees well with experimental UV-vis measurements on \SbX, which observed large absorption coefficients of $\sim$10$^5$ cm$^{-1}$ in the visible region\cite{zhou2014solution,versavel2007structural,lai2012preparation,zeng2016antimony}. Possible reasons for such high optical absorption coefficients in these indirect gap semiconductors could be attributed to their unique electronic band structures (shown in Fig. S2). On the one hand, the difference between indirect and direct gaps of \SbXsp is small (0.16 eV for \SbSsp and 0.06 eV for \SbSe) which make them still suitable for strong absorption near the band edges \cite{kondrotas2021low}. On the other hand, the relatively flat dispersions near the band extrema will lead to high DOS near the VBM and CBM and thus strong absorption \cite{kumar2014basi2}. Moreover, there is slight difference in absorption coefficients along different directions. The thickness-dependent radiative efficiencies (Fig. \ref{fig_optical}b) further show the same trend regardless of different orientations, which results from the relatively large absorption coefficients and the same value of the optical band gap along different directions. The radiative efficiencies are larger in \SbSesp than \SbSsp since the band gap of \SbSesp is closer to the optimal band gap predicted by the SQ model. For both \SbSsp and \SbSe, the radiative efficiencies approach the \ac{SQ} limit at a thickness of 2 $\mu$m. The results above indicate that absorption is not a limiting factor in the conversion efficiencies along different directions in \SbSsp and \SbSe. Other effects such as defects and interfaces should be further considered in order to improve the efficiency in \SbX.

\section{Conclusions}

The standard description of \ce{Sb2X3} in the literature refers to them as 1D semiconductors, where electrons and holes can readily diffuse \textit{along}, but not \textit{between}, ribbons in the crystal. This has lead to a focus on thin-film synthesis to achieve optimal [100] orientations. 
Our first-principles investigation has shown that the chemical binding energies between ribbons (\textgreater 10 kJ·mol$^{-1}$) falls between the \ac{vdW} and ionic/covalent regime.
Analysis of the carrier effective masses does reveal a strong anisotropy, but the behaviour is not 1D, with stronger dispersion found in [010] compared to [001].
The Fermi surfaces for electrons and holes illustrate this complexity with a combination of 3D (holes in \ce{Sb2S3}) and quasi-2D transport.  
The anisotropy carries through to the dielectric response of the crystals with much stronger screening in the \textit{ab} plane compared to along the \textit{c} axis.
However, the resulting optical absorption profiles are less sensitive and yield the same radiative limit for photovoltaic applications. 
The origin of these anisotropic effects is linked to  lone pair formation associated with the filled Sb 5\textit{s} orbitals which distorts the Sb coordination environment. 

\section*{Acknowledgements}
We are grateful to the UK Materials and Molecular Modelling Hub for computational resources, which is partially funded by EPSRC (EP/P020194/1 and EP/T022213/1). Xinwei Wang acknowledges Imperial College London for the funding of a President's PhD Scholarship. Seán R. Kavanagh acknowledges the EPSRC Centre for Doctoral Training in the Advanced Characterisation of Materials (CDT-ACM)(EP/S023259/1) for funding a PhD studentship. Alex M. Ganose was supported by EPSRC Fellowship EP/T033231/1. Xinwei Wang thanks Chengcheng Xiao and Sunghyun Kim for advice on the computational analysis.

\section*{data availability}
The data that support the findings of this study are available in an online repository at \url{https://doi.org/10.5281/zenodo.XXX}.

\section*{references}
\bibliographystyle{rsc}
\bibliography{references}

\providecommand*{\mcitethebibliography}{\thebibliography}
\csname @ifundefined\endcsname{endmcitethebibliography}
{\let\endmcitethebibliography\endthebibliography}{}
\begin{mcitethebibliography}{78}
\providecommand*{\natexlab}[1]{#1}
\providecommand*{\mciteSetBstSublistMode}[1]{}
\providecommand*{\mciteSetBstMaxWidthForm}[2]{}
\providecommand*{\mciteBstWouldAddEndPuncttrue}
  {\def\EndOfBibitem{\unskip.}}
\providecommand*{\mciteBstWouldAddEndPunctfalse}
  {\let\EndOfBibitem\relax}
\providecommand*{\mciteSetBstMidEndSepPunct}[3]{}
\providecommand*{\mciteSetBstSublistLabelBeginEnd}[3]{}
\providecommand*{\EndOfBibitem}{}
\mciteSetBstSublistMode{f}
\mciteSetBstMaxWidthForm{subitem}
{(\emph{\alph{mcitesubitemcount}})}
\mciteSetBstSublistLabelBeginEnd{\mcitemaxwidthsubitemform\space}
{\relax}{\relax}

\bibitem[Green \emph{et~al.}(2020)Green, Dunlop, Hohl-Ebinger, Yoshita,
  Kopidakis, and Ho-Baillie]{green2020solar}
M.~A. Green, E.~D. Dunlop, J.~Hohl-Ebinger, M.~Yoshita, N.~Kopidakis and A.~W.
  Ho-Baillie, \emph{Progress in Photovoltaics: Research and Applications},
  2020, \textbf{28}, 3--15\relax
\mciteBstWouldAddEndPuncttrue
\mciteSetBstMidEndSepPunct{\mcitedefaultmidpunct}
{\mcitedefaultendpunct}{\mcitedefaultseppunct}\relax
\EndOfBibitem
\bibitem[Chen \emph{et~al.}(2015)Chen, Li, Zhou, Chen, Luo, Liu, Zeng, Yang,
  Zhang, Han,\emph{et~al.}]{chen2015optical}
C.~Chen, W.~Li, Y.~Zhou, C.~Chen, M.~Luo, X.~Liu, K.~Zeng, B.~Yang, C.~Zhang,
  J.~Han \emph{et~al.}, \emph{Appl. Phys. Lett.}, 2015, \textbf{107},
  043905\relax
\mciteBstWouldAddEndPuncttrue
\mciteSetBstMidEndSepPunct{\mcitedefaultmidpunct}
{\mcitedefaultendpunct}{\mcitedefaultseppunct}\relax
\EndOfBibitem
\bibitem[Ghosh and Varma(1979)]{ghosh1979optical}
C.~Ghosh and B.~Varma, \emph{Thin solid films}, 1979, \textbf{60}, 61--65\relax
\mciteBstWouldAddEndPuncttrue
\mciteSetBstMidEndSepPunct{\mcitedefaultmidpunct}
{\mcitedefaultendpunct}{\mcitedefaultseppunct}\relax
\EndOfBibitem
\bibitem[Zhou \emph{et~al.}(2014)Zhou, Leng, Xia, Zhong, Song, Liu, Yang,
  Zhang, Chen, Zhou,\emph{et~al.}]{zhou2014solution}
Y.~Zhou, M.~Leng, Z.~Xia, J.~Zhong, H.~Song, X.~Liu, B.~Yang, J.~Zhang,
  J.~Chen, K.~Zhou \emph{et~al.}, \emph{Adv. Energy Mater.}, 2014, \textbf{4},
  1301846\relax
\mciteBstWouldAddEndPuncttrue
\mciteSetBstMidEndSepPunct{\mcitedefaultmidpunct}
{\mcitedefaultendpunct}{\mcitedefaultseppunct}\relax
\EndOfBibitem
\bibitem[Choi \emph{et~al.}(2014)Choi, Lee, Noh, Kim, and Seok]{choi2014highly}
Y.~C. Choi, D.~U. Lee, J.~H. Noh, E.~K. Kim and S.~I. Seok, \emph{Adv. Funct.
  Mater.}, 2014, \textbf{24}, 3587--3592\relax
\mciteBstWouldAddEndPuncttrue
\mciteSetBstMidEndSepPunct{\mcitedefaultmidpunct}
{\mcitedefaultendpunct}{\mcitedefaultseppunct}\relax
\EndOfBibitem
\bibitem[Li \emph{et~al.}(2019)Li, Liang, Li, Liu, Zhang, Guo, Chen, Shen, San,
  Yu,\emph{et~al.}]{li20199}
Z.~Li, X.~Liang, G.~Li, H.~Liu, H.~Zhang, J.~Guo, J.~Chen, K.~Shen, X.~San,
  W.~Yu \emph{et~al.}, \emph{Nat. Commun.}, 2019, \textbf{10}, 1--9\relax
\mciteBstWouldAddEndPuncttrue
\mciteSetBstMidEndSepPunct{\mcitedefaultmidpunct}
{\mcitedefaultendpunct}{\mcitedefaultseppunct}\relax
\EndOfBibitem
\bibitem[Shockley and Queisser(1961)]{shockley1961detailed}
W.~Shockley and H.~J. Queisser, \emph{J. Appl. Phys.}, 1961, \textbf{32},
  510--519\relax
\mciteBstWouldAddEndPuncttrue
\mciteSetBstMidEndSepPunct{\mcitedefaultmidpunct}
{\mcitedefaultendpunct}{\mcitedefaultseppunct}\relax
\EndOfBibitem
\bibitem[Kim \emph{et~al.}(2021)Kim, Ji, Jang, Jeong, Choi, Kim, Nam, and
  Shin]{kim2021importance}
J.~Kim, S.~Ji, Y.~Jang, G.~Jeong, J.~Choi, D.~Kim, S.-W. Nam and B.~Shin,
  \emph{Sol. RRL}, 2021, \textbf{5}, 2100327\relax
\mciteBstWouldAddEndPuncttrue
\mciteSetBstMidEndSepPunct{\mcitedefaultmidpunct}
{\mcitedefaultendpunct}{\mcitedefaultseppunct}\relax
\EndOfBibitem
\bibitem[Hobson \emph{et~al.}(2020)Hobson, Phillips, Hutter, Shiel, Swallow,
  Savory, Nayak, Mariotti, Das, Bowen,\emph{et~al.}]{hobson2020isotype}
T.~D. Hobson, L.~J. Phillips, O.~S. Hutter, H.~Shiel, J.~E. Swallow, C.~N.
  Savory, P.~K. Nayak, S.~Mariotti, B.~Das, L.~Bowen \emph{et~al.}, \emph{Chem.
  Mater.}, 2020, \textbf{32}, 2621--2630\relax
\mciteBstWouldAddEndPuncttrue
\mciteSetBstMidEndSepPunct{\mcitedefaultmidpunct}
{\mcitedefaultendpunct}{\mcitedefaultseppunct}\relax
\EndOfBibitem
\bibitem[Caruso \emph{et~al.}(2015)Caruso, Filip, and
  Giustino]{caruso2015excitons}
F.~Caruso, M.~R. Filip and F.~Giustino, \emph{Phys. Rev. B}, 2015, \textbf{92},
  125134\relax
\mciteBstWouldAddEndPuncttrue
\mciteSetBstMidEndSepPunct{\mcitedefaultmidpunct}
{\mcitedefaultendpunct}{\mcitedefaultseppunct}\relax
\EndOfBibitem
\bibitem[Song \emph{et~al.}(2017)Song, Li, Zhang, Zhou, Luo, Chen, Yang, Ge,
  Wu, and Tang]{song2017highly}
H.~Song, T.~Li, J.~Zhang, Y.~Zhou, J.~Luo, C.~Chen, B.~Yang, C.~Ge, Y.~Wu and
  J.~Tang, \emph{Adv. Mater.}, 2017, \textbf{29}, 1700441\relax
\mciteBstWouldAddEndPuncttrue
\mciteSetBstMidEndSepPunct{\mcitedefaultmidpunct}
{\mcitedefaultendpunct}{\mcitedefaultseppunct}\relax
\EndOfBibitem
\bibitem[Guo \emph{et~al.}(2018)Guo, Zhang, Qin, Li, Li, Qian, and
  Yan]{guo2018tunable}
L.~Guo, B.~Zhang, Y.~Qin, D.~Li, L.~Li, X.~Qian and F.~Yan, \emph{Sol. RRL},
  2018, \textbf{2}, 1800128\relax
\mciteBstWouldAddEndPuncttrue
\mciteSetBstMidEndSepPunct{\mcitedefaultmidpunct}
{\mcitedefaultendpunct}{\mcitedefaultseppunct}\relax
\EndOfBibitem
\bibitem[Yang \emph{et~al.}(2018)Yang, Ahn, Oh, Tan, Lee, Park, Kwon, Kim, Jo,
  Kim,\emph{et~al.}]{yang2018adjusting}
W.~Yang, J.~Ahn, Y.~Oh, J.~Tan, H.~Lee, J.~Park, H.-C. Kwon, J.~Kim, W.~Jo,
  J.~Kim \emph{et~al.}, \emph{Adv. Energy Mater.}, 2018, \textbf{8},
  1702888\relax
\mciteBstWouldAddEndPuncttrue
\mciteSetBstMidEndSepPunct{\mcitedefaultmidpunct}
{\mcitedefaultendpunct}{\mcitedefaultseppunct}\relax
\EndOfBibitem
\bibitem[Gusm\~ao \emph{et~al.}(2019)Gusm\~ao, Sofer, Luxa, and
  Pumera]{gusmao2019antimony}
R.~Gusm\~ao, Z.~Sofer, J.~Luxa and M.~Pumera, \emph{ACS Sustain. Chem. Eng.},
  2019, \textbf{7}, 15790--15798\relax
\mciteBstWouldAddEndPuncttrue
\mciteSetBstMidEndSepPunct{\mcitedefaultmidpunct}
{\mcitedefaultendpunct}{\mcitedefaultseppunct}\relax
\EndOfBibitem
\bibitem[Zhou \emph{et~al.}(2015)Zhou, Wang, Chen, Qin, Liu, Chen, Xue, Luo,
  Cao, Cheng,\emph{et~al.}]{zhou2015thin}
Y.~Zhou, L.~Wang, S.~Chen, S.~Qin, X.~Liu, J.~Chen, D.-J. Xue, M.~Luo, Y.~Cao,
  Y.~Cheng \emph{et~al.}, \emph{Nat. Photonics}, 2015, \textbf{9},
  409--415\relax
\mciteBstWouldAddEndPuncttrue
\mciteSetBstMidEndSepPunct{\mcitedefaultmidpunct}
{\mcitedefaultendpunct}{\mcitedefaultseppunct}\relax
\EndOfBibitem
\bibitem[Yuan \emph{et~al.}(2016)Yuan, Deng, Dong, Yang, Qiao, Hu, Song, Song,
  He, and Tang]{yuan2016efficient}
S.~Yuan, H.~Deng, D.~Dong, X.~Yang, K.~Qiao, C.~Hu, H.~Song, H.~Song, Z.~He and
  J.~Tang, \emph{Sol. Energy Mater. Sol. Cells}, 2016, \textbf{157},
  887--893\relax
\mciteBstWouldAddEndPuncttrue
\mciteSetBstMidEndSepPunct{\mcitedefaultmidpunct}
{\mcitedefaultendpunct}{\mcitedefaultseppunct}\relax
\EndOfBibitem
\bibitem[Wang \emph{et~al.}(2017)Wang, Li, Li, Chen, Deng, Gao, Zhao, Jiang,
  Li, Huang,\emph{et~al.}]{wang2017stable}
L.~Wang, D.-B. Li, K.~Li, C.~Chen, H.-X. Deng, L.~Gao, Y.~Zhao, F.~Jiang,
  L.~Li, F.~Huang \emph{et~al.}, \emph{Nat. Energy}, 2017, \textbf{2},
  1--9\relax
\mciteBstWouldAddEndPuncttrue
\mciteSetBstMidEndSepPunct{\mcitedefaultmidpunct}
{\mcitedefaultendpunct}{\mcitedefaultseppunct}\relax
\EndOfBibitem
\bibitem[Kondrotas \emph{et~al.}(2019)Kondrotas, Zhang, Wang, and
  Tang]{kondrotas2019growth}
R.~Kondrotas, J.~Zhang, C.~Wang and J.~Tang, \emph{Sol. Energy Mater. Sol.
  Cells}, 2019, \textbf{199}, 16--23\relax
\mciteBstWouldAddEndPuncttrue
\mciteSetBstMidEndSepPunct{\mcitedefaultmidpunct}
{\mcitedefaultendpunct}{\mcitedefaultseppunct}\relax
\EndOfBibitem
\bibitem[Zeng \emph{et~al.}(2020)Zeng, Sun, Huang, Nielsen, Ji, Sha, Yuan,
  Zhang, Yan, Liu,\emph{et~al.}]{zeng2020quasi}
Y.~Zeng, K.~Sun, J.~Huang, M.~P. Nielsen, F.~Ji, C.~Sha, S.~Yuan, X.~Zhang,
  C.~Yan, X.~Liu \emph{et~al.}, \emph{ACS Appl. Mater. Interfaces}, 2020,
  \textbf{12}, 22825--22834\relax
\mciteBstWouldAddEndPuncttrue
\mciteSetBstMidEndSepPunct{\mcitedefaultmidpunct}
{\mcitedefaultendpunct}{\mcitedefaultseppunct}\relax
\EndOfBibitem
\bibitem[McKenna(2021)]{mckenna2021self}
K.~P. McKenna, \emph{Adv. Electron. Mater}, 2021, \textbf{7}, 2000908\relax
\mciteBstWouldAddEndPuncttrue
\mciteSetBstMidEndSepPunct{\mcitedefaultmidpunct}
{\mcitedefaultendpunct}{\mcitedefaultseppunct}\relax
\EndOfBibitem
\bibitem[Kohn and Sham(1965)]{kohn1965self}
W.~Kohn and L.~J. Sham, \emph{Phys. Rev.}, 1965, \textbf{140}, A1133\relax
\mciteBstWouldAddEndPuncttrue
\mciteSetBstMidEndSepPunct{\mcitedefaultmidpunct}
{\mcitedefaultendpunct}{\mcitedefaultseppunct}\relax
\EndOfBibitem
\bibitem[Dreizler and Gross(1990)]{dreizler1990density}
R.~M. Dreizler and E.~K. Gross, \emph{Density Functional Theory}, Springer,
  1990, pp. 245--271\relax
\mciteBstWouldAddEndPuncttrue
\mciteSetBstMidEndSepPunct{\mcitedefaultmidpunct}
{\mcitedefaultendpunct}{\mcitedefaultseppunct}\relax
\EndOfBibitem
\bibitem[Kresse and Furthm{\"u}ller(1996)]{kresse1996efficient}
G.~Kresse and J.~Furthm{\"u}ller, \emph{Phys. Rev. B}, 1996, \textbf{54},
  11169\relax
\mciteBstWouldAddEndPuncttrue
\mciteSetBstMidEndSepPunct{\mcitedefaultmidpunct}
{\mcitedefaultendpunct}{\mcitedefaultseppunct}\relax
\EndOfBibitem
\bibitem[Kresse and Joubert(1999)]{kresse1999ultrasoft}
G.~Kresse and D.~Joubert, \emph{Phys. Rev. B}, 1999, \textbf{59}, 1758\relax
\mciteBstWouldAddEndPuncttrue
\mciteSetBstMidEndSepPunct{\mcitedefaultmidpunct}
{\mcitedefaultendpunct}{\mcitedefaultseppunct}\relax
\EndOfBibitem
\bibitem[Heyd \emph{et~al.}(2003)Heyd, Scuseria, and Ernzerhof]{heyd2003hybrid}
J.~Heyd, G.~E. Scuseria and M.~Ernzerhof, \emph{J. Chem. Phys.}, 2003,
  \textbf{118}, 8207--8215\relax
\mciteBstWouldAddEndPuncttrue
\mciteSetBstMidEndSepPunct{\mcitedefaultmidpunct}
{\mcitedefaultendpunct}{\mcitedefaultseppunct}\relax
\EndOfBibitem
\bibitem[Krukau \emph{et~al.}(2006)Krukau, Vydrov, Izmaylov, and
  Scuseria]{krukau2006influence}
A.~V. Krukau, O.~A. Vydrov, A.~F. Izmaylov and G.~E. Scuseria, \emph{J. Chem.
  Phys.}, 2006, \textbf{125}, 224106\relax
\mciteBstWouldAddEndPuncttrue
\mciteSetBstMidEndSepPunct{\mcitedefaultmidpunct}
{\mcitedefaultendpunct}{\mcitedefaultseppunct}\relax
\EndOfBibitem
\bibitem[Kavanagh \emph{et~al.}(2021)Kavanagh, Walsh, and
  Scanlon]{kavanagh2021rapid}
S.~R. Kavanagh, A.~Walsh and D.~O. Scanlon, \emph{ACS Energy Lett.}, 2021,
  \textbf{6}, 1392--1398\relax
\mciteBstWouldAddEndPuncttrue
\mciteSetBstMidEndSepPunct{\mcitedefaultmidpunct}
{\mcitedefaultendpunct}{\mcitedefaultseppunct}\relax
\EndOfBibitem
\bibitem[Klime{\v{s}} \emph{et~al.}(2011)Klime{\v{s}}, Bowler, and
  Michaelides]{klimevs2011van}
J.~Klime{\v{s}}, D.~R. Bowler and A.~Michaelides, \emph{Phys. Rev. B}, 2011,
  \textbf{83}, 195131\relax
\mciteBstWouldAddEndPuncttrue
\mciteSetBstMidEndSepPunct{\mcitedefaultmidpunct}
{\mcitedefaultendpunct}{\mcitedefaultseppunct}\relax
\EndOfBibitem
\bibitem[Grimme(2004)]{grimme2004accurate}
S.~Grimme, \emph{J. Comput. Chem.}, 2004, \textbf{25}, 1463--1473\relax
\mciteBstWouldAddEndPuncttrue
\mciteSetBstMidEndSepPunct{\mcitedefaultmidpunct}
{\mcitedefaultendpunct}{\mcitedefaultseppunct}\relax
\EndOfBibitem
\bibitem[Savory and Scanlon(2019)]{savory2019complex}
C.~N. Savory and D.~O. Scanlon, \emph{J. Mater. Chem. A}, 2019, \textbf{7},
  10739--10744\relax
\mciteBstWouldAddEndPuncttrue
\mciteSetBstMidEndSepPunct{\mcitedefaultmidpunct}
{\mcitedefaultendpunct}{\mcitedefaultseppunct}\relax
\EndOfBibitem
\bibitem[cry()]{crystalmaker}
\emph{CrystalMaker, {CrystalMaker Software Ltd, Oxford, England},
  (www.crystalmaker.com)}\relax
\mciteBstWouldAddEndPuncttrue
\mciteSetBstMidEndSepPunct{\mcitedefaultmidpunct}
{\mcitedefaultendpunct}{\mcitedefaultseppunct}\relax
\EndOfBibitem
\bibitem[Ganose \emph{et~al.}(2018)Ganose, Jackson, and
  Scanlon]{ganose2018sumo}
A.~M. Ganose, A.~J. Jackson and D.~O. Scanlon, \emph{J. Open Source Softw.},
  2018, \textbf{3}, 717\relax
\mciteBstWouldAddEndPuncttrue
\mciteSetBstMidEndSepPunct{\mcitedefaultmidpunct}
{\mcitedefaultendpunct}{\mcitedefaultseppunct}\relax
\EndOfBibitem
\bibitem[Dronskowski and Bl{\"o}chl(1993)]{dronskowski1993crystal}
R.~Dronskowski and P.~E. Bl{\"o}chl, \emph{J. Phys. Chem.}, 1993, \textbf{97},
  8617--8624\relax
\mciteBstWouldAddEndPuncttrue
\mciteSetBstMidEndSepPunct{\mcitedefaultmidpunct}
{\mcitedefaultendpunct}{\mcitedefaultseppunct}\relax
\EndOfBibitem
\bibitem[Ganose \emph{et~al.}(2021)Ganose, Searle, Jain, and
  Griffin]{ganose2021ifermi}
A.~M. Ganose, A.~Searle, A.~Jain and S.~M. Griffin, \emph{J. Open Source
  Softw.}, 2021, \textbf{6}, 3089\relax
\mciteBstWouldAddEndPuncttrue
\mciteSetBstMidEndSepPunct{\mcitedefaultmidpunct}
{\mcitedefaultendpunct}{\mcitedefaultseppunct}\relax
\EndOfBibitem
\bibitem[Ganose \emph{et~al.}(2021)Ganose, Park, Faghaninia, Woods-Robinson,
  Persson, and Jain]{ganose2021efficient}
A.~M. Ganose, J.~Park, A.~Faghaninia, R.~Woods-Robinson, K.~A. Persson and
  A.~Jain, \emph{Nat. Commun.}, 2021, \textbf{12}, 1--9\relax
\mciteBstWouldAddEndPuncttrue
\mciteSetBstMidEndSepPunct{\mcitedefaultmidpunct}
{\mcitedefaultendpunct}{\mcitedefaultseppunct}\relax
\EndOfBibitem
\bibitem[Ashcroft \emph{et~al.}(1976)Ashcroft,
  Mermin,\emph{et~al.}]{ashcroft1976solid}
N.~W. Ashcroft, N.~D. Mermin \emph{et~al.}, \emph{Solid State Physics},
  1976\relax
\mciteBstWouldAddEndPuncttrue
\mciteSetBstMidEndSepPunct{\mcitedefaultmidpunct}
{\mcitedefaultendpunct}{\mcitedefaultseppunct}\relax
\EndOfBibitem
\bibitem[Madsen and Singh(2006)]{madsen2006boltztrap}
G.~K. Madsen and D.~J. Singh, \emph{Comput. Phys. Commun.}, 2006, \textbf{175},
  67--71\relax
\mciteBstWouldAddEndPuncttrue
\mciteSetBstMidEndSepPunct{\mcitedefaultmidpunct}
{\mcitedefaultendpunct}{\mcitedefaultseppunct}\relax
\EndOfBibitem
\bibitem[Gibbs \emph{et~al.}(2017)Gibbs, Ricci, Li, Zhu, Persson, Ceder,
  Hautier, Jain, and Snyder]{gibbs2017effective}
Z.~M. Gibbs, F.~Ricci, G.~Li, H.~Zhu, K.~Persson, G.~Ceder, G.~Hautier, A.~Jain
  and G.~J. Snyder, \emph{NPJ Comput. Mater.}, 2017, \textbf{3}, 1--7\relax
\mciteBstWouldAddEndPuncttrue
\mciteSetBstMidEndSepPunct{\mcitedefaultmidpunct}
{\mcitedefaultendpunct}{\mcitedefaultseppunct}\relax
\EndOfBibitem
\bibitem[Gajdo{\v{s}} \emph{et~al.}(2006)Gajdo{\v{s}}, Hummer, Kresse,
  Furthm{\"u}ller, and Bechstedt]{gajdovs2006linear}
M.~Gajdo{\v{s}}, K.~Hummer, G.~Kresse, J.~Furthm{\"u}ller and F.~Bechstedt,
  \emph{Phys. Rev. B}, 2006, \textbf{73}, 045112\relax
\mciteBstWouldAddEndPuncttrue
\mciteSetBstMidEndSepPunct{\mcitedefaultmidpunct}
{\mcitedefaultendpunct}{\mcitedefaultseppunct}\relax
\EndOfBibitem
\bibitem[Hunter(2007)]{Hunter:2007}
J.~D. Hunter, \emph{Comput. Sci. Eng.}, 2007, \textbf{9}, 90--95\relax
\mciteBstWouldAddEndPuncttrue
\mciteSetBstMidEndSepPunct{\mcitedefaultmidpunct}
{\mcitedefaultendpunct}{\mcitedefaultseppunct}\relax
\EndOfBibitem
\bibitem[Kyono \emph{et~al.}(2002)Kyono, Kimata, Matsuhisa, Miyashita, and
  Okamoto]{kyono2002low}
A.~Kyono, M.~Kimata, M.~Matsuhisa, Y.~Miyashita and K.~Okamoto, \emph{Phys.
  Chem. Miner.}, 2002, \textbf{29}, 254--260\relax
\mciteBstWouldAddEndPuncttrue
\mciteSetBstMidEndSepPunct{\mcitedefaultmidpunct}
{\mcitedefaultendpunct}{\mcitedefaultseppunct}\relax
\EndOfBibitem
\bibitem[Savadogo and Mandal(1992)]{savadogo1992studies}
O.~Savadogo and K.~Mandal, \emph{Sol. Energy Mater. Sol. Cells}, 1992,
  \textbf{26}, 117--136\relax
\mciteBstWouldAddEndPuncttrue
\mciteSetBstMidEndSepPunct{\mcitedefaultmidpunct}
{\mcitedefaultendpunct}{\mcitedefaultseppunct}\relax
\EndOfBibitem
\bibitem[Salem and Selim(2001)]{salem2001structure}
A.~Salem and M.~S. Selim, \emph{J. Phys. D}, 2001, \textbf{34}, 12\relax
\mciteBstWouldAddEndPuncttrue
\mciteSetBstMidEndSepPunct{\mcitedefaultmidpunct}
{\mcitedefaultendpunct}{\mcitedefaultseppunct}\relax
\EndOfBibitem
\bibitem[Efthimiopoulos \emph{et~al.}(2013)Efthimiopoulos, Zhang, Kucway, Park,
  Ewing, and Wang]{efthimiopoulos2013sb}
I.~Efthimiopoulos, J.~Zhang, M.~Kucway, C.~Park, R.~C. Ewing and Y.~Wang,
  \emph{Sci. Rep.}, 2013, \textbf{3}, 1--8\relax
\mciteBstWouldAddEndPuncttrue
\mciteSetBstMidEndSepPunct{\mcitedefaultmidpunct}
{\mcitedefaultendpunct}{\mcitedefaultseppunct}\relax
\EndOfBibitem
\bibitem[Voutsas \emph{et~al.}(1985)Voutsas, Papazoglou, Rentzeperis, and
  Siapkas]{voutsas1985crystal}
G.~Voutsas, A.~Papazoglou, P.~Rentzeperis and D.~Siapkas, \emph{Z. Kristallogr.
  Cryst. Mater.}, 1985, \textbf{171}, 261--268\relax
\mciteBstWouldAddEndPuncttrue
\mciteSetBstMidEndSepPunct{\mcitedefaultmidpunct}
{\mcitedefaultendpunct}{\mcitedefaultseppunct}\relax
\EndOfBibitem
\bibitem[Hurych \emph{et~al.}(1974)Hurych, Davis, Buczek, Wood, Lapeyre, and
  Baer]{hurych1974photoemission}
Z.~Hurych, D.~Davis, D.~Buczek, C.~Wood, G.~Lapeyre and A.~Baer, \emph{Phys.
  Rev. B}, 1974, \textbf{9}, 4392\relax
\mciteBstWouldAddEndPuncttrue
\mciteSetBstMidEndSepPunct{\mcitedefaultmidpunct}
{\mcitedefaultendpunct}{\mcitedefaultseppunct}\relax
\EndOfBibitem
\bibitem[Stoliaroff \emph{et~al.}(2020)Stoliaroff, Lecomte, Rubel, Jobic,
  Zhang, Latouche, and Rocquefelte]{stoliaroff2020deciphering}
A.~Stoliaroff, A.~Lecomte, O.~Rubel, S.~Jobic, X.~Zhang, C.~Latouche and
  X.~Rocquefelte, \emph{ACS Appl. Energy Mater.}, 2020, \textbf{3},
  2496--2509\relax
\mciteBstWouldAddEndPuncttrue
\mciteSetBstMidEndSepPunct{\mcitedefaultmidpunct}
{\mcitedefaultendpunct}{\mcitedefaultseppunct}\relax
\EndOfBibitem
\bibitem[Deringer \emph{et~al.}(2015)Deringer, Stoffel, Wuttig, and
  Dronskowski]{deringer2015vibrational}
V.~L. Deringer, R.~P. Stoffel, M.~Wuttig and R.~Dronskowski, \emph{Chem. Sci},
  2015, \textbf{6}, 5255--5262\relax
\mciteBstWouldAddEndPuncttrue
\mciteSetBstMidEndSepPunct{\mcitedefaultmidpunct}
{\mcitedefaultendpunct}{\mcitedefaultseppunct}\relax
\EndOfBibitem
\bibitem[Mounet \emph{et~al.}(2018)Mounet, Gibertini, Schwaller, Campi, Merkys,
  Marrazzo, Sohier, Castelli, Cepellotti, Pizzi,\emph{et~al.}]{mounet2018two}
N.~Mounet, M.~Gibertini, P.~Schwaller, D.~Campi, A.~Merkys, A.~Marrazzo,
  T.~Sohier, I.~E. Castelli, A.~Cepellotti, G.~Pizzi \emph{et~al.}, \emph{Nat.
  Nanotechnol.}, 2018, \textbf{13}, 246--252\relax
\mciteBstWouldAddEndPuncttrue
\mciteSetBstMidEndSepPunct{\mcitedefaultmidpunct}
{\mcitedefaultendpunct}{\mcitedefaultseppunct}\relax
\EndOfBibitem
\bibitem[Bj{\"o}rkman \emph{et~al.}(2012)Bj{\"o}rkman, Gulans, Krasheninnikov,
  and Nieminen]{bjorkman2012van}
T.~Bj{\"o}rkman, A.~Gulans, A.~V. Krasheninnikov and R.~M. Nieminen,
  \emph{Phys. Rev. Lett.}, 2012, \textbf{108}, 235502\relax
\mciteBstWouldAddEndPuncttrue
\mciteSetBstMidEndSepPunct{\mcitedefaultmidpunct}
{\mcitedefaultendpunct}{\mcitedefaultseppunct}\relax
\EndOfBibitem
\bibitem[Garrett and Grisham(1999)]{garrett1999biochemistry}
R.~H. Garrett and C.~M. Grisham, \emph{Biochemistry}, 1999\relax
\mciteBstWouldAddEndPuncttrue
\mciteSetBstMidEndSepPunct{\mcitedefaultmidpunct}
{\mcitedefaultendpunct}{\mcitedefaultseppunct}\relax
\EndOfBibitem
\bibitem[Filip \emph{et~al.}(2013)Filip, Patrick, and Giustino]{filip2013g}
M.~R. Filip, C.~E. Patrick and F.~Giustino, \emph{Phys. Rev. B}, 2013,
  \textbf{87}, 205125\relax
\mciteBstWouldAddEndPuncttrue
\mciteSetBstMidEndSepPunct{\mcitedefaultmidpunct}
{\mcitedefaultendpunct}{\mcitedefaultseppunct}\relax
\EndOfBibitem
\bibitem[Caracas and Gonze(2005)]{caracas2005first}
R.~Caracas and X.~Gonze, \emph{Phys. Chem. Miner.}, 2005, \textbf{32},
  295--300\relax
\mciteBstWouldAddEndPuncttrue
\mciteSetBstMidEndSepPunct{\mcitedefaultmidpunct}
{\mcitedefaultendpunct}{\mcitedefaultseppunct}\relax
\EndOfBibitem
\bibitem[Ko{\c{c}} \emph{et~al.}(2012)Ko{\c{c}}, Mamedov, Deligoz, and
  Ozisik]{kocc2012first}
H.~Ko{\c{c}}, A.~M. Mamedov, E.~Deligoz and H.~Ozisik, \emph{Solid State Sci.},
  2012, \textbf{14}, 1211--1220\relax
\mciteBstWouldAddEndPuncttrue
\mciteSetBstMidEndSepPunct{\mcitedefaultmidpunct}
{\mcitedefaultendpunct}{\mcitedefaultseppunct}\relax
\EndOfBibitem
\bibitem[Radzwan \emph{et~al.}(2017)Radzwan, Ahmed, Shaari, Lawal, and
  Ng]{radzwan2017first}
A.~Radzwan, R.~Ahmed, A.~Shaari, A.~Lawal and Y.~X. Ng, \emph{Malays. J.
  Fundam. Appl. Sci.}, 2017, \textbf{13}, 285--289\relax
\mciteBstWouldAddEndPuncttrue
\mciteSetBstMidEndSepPunct{\mcitedefaultmidpunct}
{\mcitedefaultendpunct}{\mcitedefaultseppunct}\relax
\EndOfBibitem
\bibitem[Ganose \emph{et~al.}(2016)Ganose, Butler, Walsh, and
  Scanlon]{ganose2016relativistic}
A.~M. Ganose, K.~T. Butler, A.~Walsh and D.~O. Scanlon, \emph{J. Mater. Chem.
  A}, 2016, \textbf{4}, 2060--2068\relax
\mciteBstWouldAddEndPuncttrue
\mciteSetBstMidEndSepPunct{\mcitedefaultmidpunct}
{\mcitedefaultendpunct}{\mcitedefaultseppunct}\relax
\EndOfBibitem
\bibitem[Walsh \emph{et~al.}(2011)Walsh, Payne, Egdell, and
  Watson]{walsh2011stereochemistry}
A.~Walsh, D.~J. Payne, R.~G. Egdell and G.~W. Watson, \emph{Chem. Soc. Rev.},
  2011, \textbf{40}, 4455--4463\relax
\mciteBstWouldAddEndPuncttrue
\mciteSetBstMidEndSepPunct{\mcitedefaultmidpunct}
{\mcitedefaultendpunct}{\mcitedefaultseppunct}\relax
\EndOfBibitem
\bibitem[Walsh and Watson(2005)]{walsh2005origin}
A.~Walsh and G.~W. Watson, \emph{J. Solid State Chem.}, 2005, \textbf{178},
  1422--1428\relax
\mciteBstWouldAddEndPuncttrue
\mciteSetBstMidEndSepPunct{\mcitedefaultmidpunct}
{\mcitedefaultendpunct}{\mcitedefaultseppunct}\relax
\EndOfBibitem
\bibitem[Allen \emph{et~al.}(2013)Allen, Carey, Walsh, Scanlon, and
  Watson]{allen2013electronic}
J.~P. Allen, J.~J. Carey, A.~Walsh, D.~O. Scanlon and G.~W. Watson, \emph{J.
  Phys. Chem. C}, 2013, \textbf{117}, 14759--14769\relax
\mciteBstWouldAddEndPuncttrue
\mciteSetBstMidEndSepPunct{\mcitedefaultmidpunct}
{\mcitedefaultendpunct}{\mcitedefaultseppunct}\relax
\EndOfBibitem
\bibitem[Kavanagh \emph{et~al.}(2021)Kavanagh, Savory, Scanlon, and
  Walsh]{kavanagh2021hidden}
S.~R. Kavanagh, C.~N. Savory, D.~O. Scanlon and A.~Walsh, \emph{Mater. Horiz.},
  2021\relax
\mciteBstWouldAddEndPuncttrue
\mciteSetBstMidEndSepPunct{\mcitedefaultmidpunct}
{\mcitedefaultendpunct}{\mcitedefaultseppunct}\relax
\EndOfBibitem
\bibitem[Ganose \emph{et~al.}(2018)Ganose, Matsumoto, Buckeridge, and
  Scanlon]{ganose2018defect}
A.~M. Ganose, S.~Matsumoto, J.~Buckeridge and D.~O. Scanlon, \emph{Chem.
  Mater.}, 2018, \textbf{30}, 3827--3835\relax
\mciteBstWouldAddEndPuncttrue
\mciteSetBstMidEndSepPunct{\mcitedefaultmidpunct}
{\mcitedefaultendpunct}{\mcitedefaultseppunct}\relax
\EndOfBibitem
\bibitem[Nasr \emph{et~al.}(2016)Nasr, Maghraoui-Meherzi, and
  Kamoun-Turki]{nasr2016first}
T.~B. Nasr, H.~Maghraoui-Meherzi and N.~Kamoun-Turki, \emph{J. Alloys Compd.},
  2016, \textbf{663}, 123--127\relax
\mciteBstWouldAddEndPuncttrue
\mciteSetBstMidEndSepPunct{\mcitedefaultmidpunct}
{\mcitedefaultendpunct}{\mcitedefaultseppunct}\relax
\EndOfBibitem
\bibitem[Qiu \emph{et~al.}(2019)Qiu, Zhang, Cheng, Zheng, Yu, Jia, and
  Wu]{qiu2019crystal}
W.~Qiu, C.~Zhang, S.~Cheng, Q.~Zheng, X.~Yu, H.~Jia and B.~Wu, \emph{J. Solid
  State Chem.}, 2019, \textbf{271}, 339--345\relax
\mciteBstWouldAddEndPuncttrue
\mciteSetBstMidEndSepPunct{\mcitedefaultmidpunct}
{\mcitedefaultendpunct}{\mcitedefaultseppunct}\relax
\EndOfBibitem
\bibitem[Whalley \emph{et~al.}(2019)Whalley, Frost, Morgan, and
  Walsh]{whalley2019impact}
L.~D. Whalley, J.~M. Frost, B.~J. Morgan and A.~Walsh, \emph{Phys. Rev. B},
  2019, \textbf{99}, 085207\relax
\mciteBstWouldAddEndPuncttrue
\mciteSetBstMidEndSepPunct{\mcitedefaultmidpunct}
{\mcitedefaultendpunct}{\mcitedefaultseppunct}\relax
\EndOfBibitem
\bibitem[Albert(2015)]{albert2015torque}
S.~G. Albert, \emph{PhD thesis}, Technische Universit{\"a}t M{\"u}nchen,
  2015\relax
\mciteBstWouldAddEndPuncttrue
\mciteSetBstMidEndSepPunct{\mcitedefaultmidpunct}
{\mcitedefaultendpunct}{\mcitedefaultseppunct}\relax
\EndOfBibitem
\bibitem[Zeng \emph{et~al.}(2016)Zeng, Xue, and Tang]{zeng2016antimony}
K.~Zeng, D.-J. Xue and J.~Tang, \emph{Semicond. Sci. Technol.}, 2016,
  \textbf{31}, 063001\relax
\mciteBstWouldAddEndPuncttrue
\mciteSetBstMidEndSepPunct{\mcitedefaultmidpunct}
{\mcitedefaultendpunct}{\mcitedefaultseppunct}\relax
\EndOfBibitem
\bibitem[Walsh and Zunger(2017)]{walsh2017instilling}
A.~Walsh and A.~Zunger, \emph{Nat. Mater.}, 2017, \textbf{16}, 964--967\relax
\mciteBstWouldAddEndPuncttrue
\mciteSetBstMidEndSepPunct{\mcitedefaultmidpunct}
{\mcitedefaultendpunct}{\mcitedefaultseppunct}\relax
\EndOfBibitem
\bibitem[Liu \emph{et~al.}(2014)Liu, Chua, Sum, and Gan]{liu2014first}
Y.~Liu, K.~T.~E. Chua, T.~C. Sum and C.~K. Gan, \emph{Phys. Chem. Chem. Phys.},
  2014, \textbf{16}, 345--350\relax
\mciteBstWouldAddEndPuncttrue
\mciteSetBstMidEndSepPunct{\mcitedefaultmidpunct}
{\mcitedefaultendpunct}{\mcitedefaultseppunct}\relax
\EndOfBibitem
\bibitem[Cheng \emph{et~al.}(2019)Cheng, Cojocaru-Mir{\'e}din, Keutgen, Yu,
  K{\"u}pers, Schumacher, Golub, Raty, Dronskowski, and
  Wuttig]{cheng2019understanding}
Y.~Cheng, O.~Cojocaru-Mir{\'e}din, J.~Keutgen, Y.~Yu, M.~K{\"u}pers,
  M.~Schumacher, P.~Golub, J.-Y. Raty, R.~Dronskowski and M.~Wuttig, \emph{Adv.
  Mater.}, 2019, \textbf{31}, 1904316\relax
\mciteBstWouldAddEndPuncttrue
\mciteSetBstMidEndSepPunct{\mcitedefaultmidpunct}
{\mcitedefaultendpunct}{\mcitedefaultseppunct}\relax
\EndOfBibitem
\bibitem[Schubert \emph{et~al.}(2004)Schubert, Hofmann, Herzinger, and
  Dollase]{schubert2004generalized}
M.~Schubert, T.~Hofmann, C.~Herzinger and W.~Dollase, \emph{Thin Solid Films},
  2004, \textbf{455}, 619--623\relax
\mciteBstWouldAddEndPuncttrue
\mciteSetBstMidEndSepPunct{\mcitedefaultmidpunct}
{\mcitedefaultendpunct}{\mcitedefaultseppunct}\relax
\EndOfBibitem
\bibitem[Nasr \emph{et~al.}(2011)Nasr, Maghraoui-Meherzi, Abdallah, and
  Bennaceur]{nasr2011electronic}
T.~B. Nasr, H.~Maghraoui-Meherzi, H.~B. Abdallah and R.~Bennaceur,
  \emph{Physica B Condens. Matter.}, 2011, \textbf{406}, 287--292\relax
\mciteBstWouldAddEndPuncttrue
\mciteSetBstMidEndSepPunct{\mcitedefaultmidpunct}
{\mcitedefaultendpunct}{\mcitedefaultseppunct}\relax
\EndOfBibitem
\bibitem[Lakhdar \emph{et~al.}(2014)Lakhdar, Ouni, and
  Amlouk]{lakhdar2014dielectric}
M.~H. Lakhdar, B.~Ouni and M.~Amlouk, \emph{Mater. Sci. Semicond. Process.},
  2014, \textbf{19}, 32--39\relax
\mciteBstWouldAddEndPuncttrue
\mciteSetBstMidEndSepPunct{\mcitedefaultmidpunct}
{\mcitedefaultendpunct}{\mcitedefaultseppunct}\relax
\EndOfBibitem
\bibitem[Maghraoui-Meherzi \emph{et~al.}(2013)Maghraoui-Meherzi, Nasr, and
  Dachraoui]{maghraoui2013synthesis}
H.~Maghraoui-Meherzi, T.~B. Nasr and M.~Dachraoui, \emph{Mater. Sci. Semicond.
  Process.}, 2013, \textbf{16}, 179--184\relax
\mciteBstWouldAddEndPuncttrue
\mciteSetBstMidEndSepPunct{\mcitedefaultmidpunct}
{\mcitedefaultendpunct}{\mcitedefaultseppunct}\relax
\EndOfBibitem
\bibitem[Lawal \emph{et~al.}(2018)Lawal, Shaari, Ahmed, and
  Taura]{lawal2018investigation}
A.~Lawal, A.~Shaari, R.~Ahmed and L.~Taura, \emph{Curr. Appl. Phys.}, 2018,
  \textbf{18}, 567--575\relax
\mciteBstWouldAddEndPuncttrue
\mciteSetBstMidEndSepPunct{\mcitedefaultmidpunct}
{\mcitedefaultendpunct}{\mcitedefaultseppunct}\relax
\EndOfBibitem
\bibitem[Versavel and Haber(2007)]{versavel2007structural}
M.~Y. Versavel and J.~A. Haber, \emph{Thin Solid Films}, 2007, \textbf{515},
  7171--7176\relax
\mciteBstWouldAddEndPuncttrue
\mciteSetBstMidEndSepPunct{\mcitedefaultmidpunct}
{\mcitedefaultendpunct}{\mcitedefaultseppunct}\relax
\EndOfBibitem
\bibitem[Lai \emph{et~al.}(2012)Lai, Chen, Han, Jiang, Liu, Li, and
  Liu]{lai2012preparation}
Y.~Lai, Z.~Chen, C.~Han, L.~Jiang, F.~Liu, J.~Li and Y.~Liu, \emph{Appl. Surf.
  Sci.}, 2012, \textbf{261}, 510--514\relax
\mciteBstWouldAddEndPuncttrue
\mciteSetBstMidEndSepPunct{\mcitedefaultmidpunct}
{\mcitedefaultendpunct}{\mcitedefaultseppunct}\relax
\EndOfBibitem
\bibitem[Kondrotas \emph{et~al.}(2021)Kondrotas, Chen, Liu, Yang, and
  Tang]{kondrotas2021low}
R.~Kondrotas, C.~Chen, X.~Liu, B.~Yang and J.~Tang, \emph{J. Semicond.}, 2021,
  \textbf{42}, 031701\relax
\mciteBstWouldAddEndPuncttrue
\mciteSetBstMidEndSepPunct{\mcitedefaultmidpunct}
{\mcitedefaultendpunct}{\mcitedefaultseppunct}\relax
\EndOfBibitem
\bibitem[Kumar \emph{et~al.}(2014)Kumar, Umezawa, and Imai]{kumar2014basi2}
M.~Kumar, N.~Umezawa and M.~Imai, \emph{Appl. Phys. Express}, 2014, \textbf{7},
  071203\relax
\mciteBstWouldAddEndPuncttrue
\mciteSetBstMidEndSepPunct{\mcitedefaultmidpunct}
{\mcitedefaultendpunct}{\mcitedefaultseppunct}\relax
\EndOfBibitem
\end{mcitethebibliography}

\end{document}


\centerline{\huge \textbf{Supporting Information}}

\beginsupplement

~\\
~\\
\section{Binding energies}

\begin{figure}[h]
    \centering
    {\includegraphics[width=\textwidth]{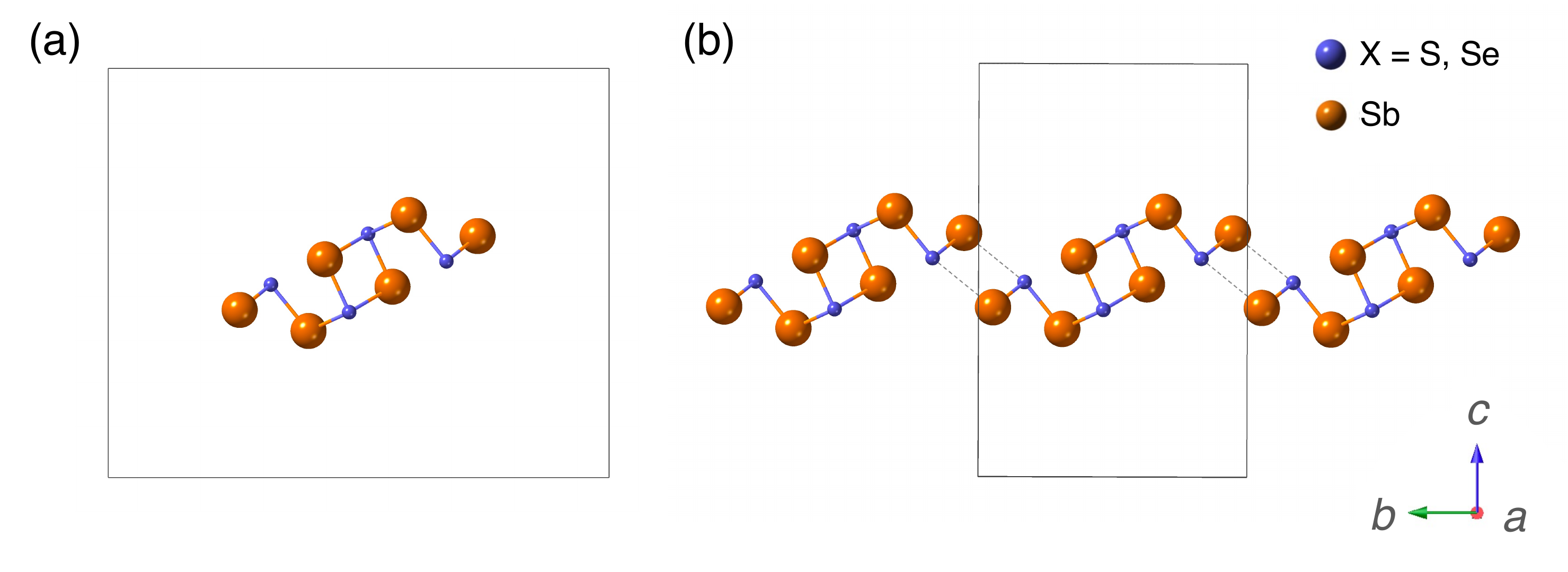}} \\
    \caption{(a) 1D and (b) 2D substructures of \ce{Sb2X3} used to calculate binding energies. The unit cells are represented by rectangles.}
    \label{fig_Ed}
\end{figure}

The substructures for calculating the inter-ribbon binding energies ($E_\textrm{b}$) are shown in Fig. \ref{fig_Ed}. Fig. \ref{fig_Ed}a is the 1D [Sb$_4$X$_6$]$_n$ ribbon in isolation and Fig. \ref{fig_Ed}b is the 2D ribbon periodically repeated along \textit{a} and \textit{b} directions. Table \ref{tab_binding_novdw} shows $E_\textrm{b}$ calculated by PBE and HSE06 functionals without vdW corrections. Compared with the results obtained by the HSE06 functional and D3 dispersion correction (shown in Table 2), the $E_\textrm{b}$ here are much lower especially along the \textit{c} direction, which is attributed to larger lattice parameters in the \textit{c} direction without vdW corrections. This indicates that the inclusion of vdW corrections is necessary to reproduce the crystal structures of \SbX.

\begin{table}[h]
\begin{tabular}{ccccc}
\hline
\multirow{2}{*}{System} & \multirow{2}{*}{Functional} & $E_{\textrm{b}(\textrm{b})}$        & \multicolumn{2}{c}{$E_{\textrm{b}(\textrm{c})}$}            \\ \cline{3-5} 
                 & & per bond & per bond (N=16) & per atom (N=20) \\ \hline
\multirow{2}{*}{Sb$_2$S$_3$} &PBE & 19.66    & 4.32           & 3.46           \\ 
&HSE06 &13.77 &3.81 &3.05 \\ [1ex]
\multirow{2}{*}{Sb$_2$Se$_3$} &PBE                & 23.68    & 3.93           & 3.14           \\ 
&HSE06 &14.87 &3.93 &3.14 \\\hline
\end{tabular}
\caption{Inter-ribbon binding energies (kJ·mol$^{-1}$) of \SbSsp and \SbSesp without vdW corrections}
\label{tab_binding_novdw}
\end{table}

\section{Electronic band structures}

\begin{figure}[h]
    \centering
    {\includegraphics[width=0.8\textwidth]{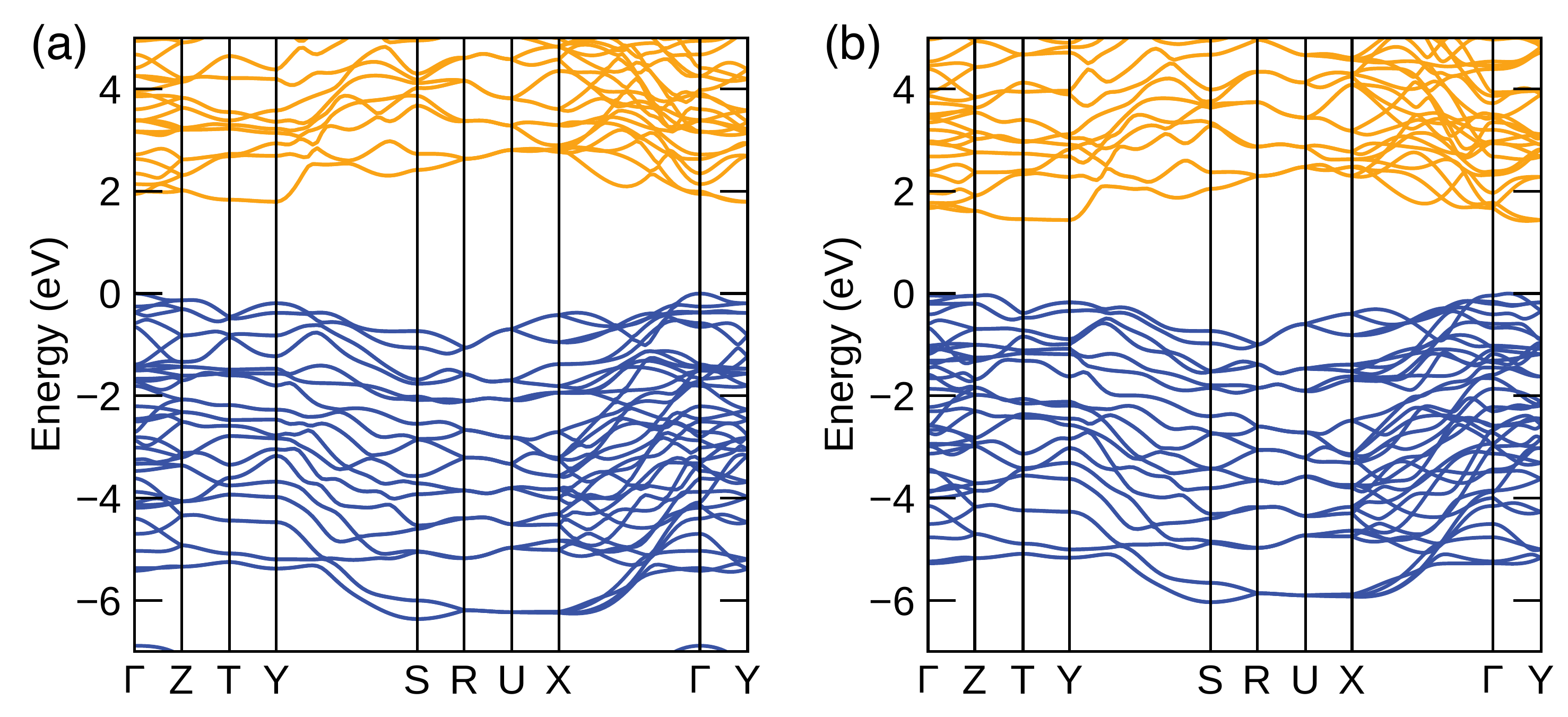}} \\
    \caption{Electronic band structures of (a) \SbSsp and (b) \SbSe.}
    \label{fig_band}
\end{figure}

Calculated electronic band structures are shown in Fig. \ref{fig_band}. According to our results, \SbSsp and \SbSesp are indirect semiconductors with indirect (direct) band gaps of 1.79 (1.95) and 1.42 (1.48) eV, respectively, which are in reasonable agreement with previous experimental \cite{yesugade1995structural,el1998substrate,versavel2007structural,liu2016green,torane1999preparation,messina2009antimony,lai2012preparation,chen2015optical} and theoretical studies \cite{vadapoo2011self,vadapoo2011electronic,caracas2005first,nasr2011electronic,kocc2012first,savory2019complex}. 

\section{Fermi surfaces}

\begin{figure}[h]
    \centering
    {\includegraphics[width=\textwidth]{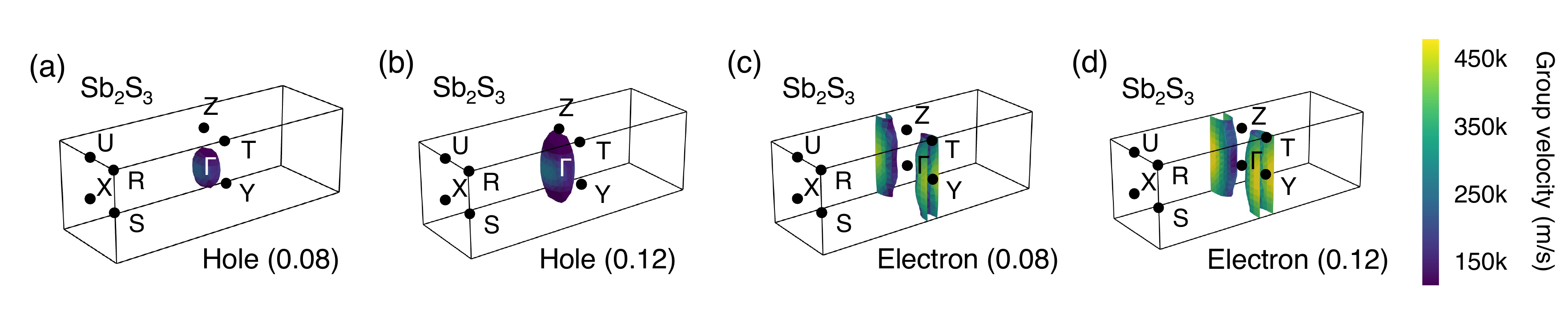}} \\
    \caption{Fermi surfaces of \SbS. (a) and (b) are hole Fermi surfaces (0.08 and 0.12 eV below the VBM, respectively), while (c) and (d) are electron Fermi surfaces (0.08 and 0.12 eV above the CBM, respectively). The different colors represent the magnitude of group velocity (m/s) }
    \label{fig_Fermi_S}
\end{figure}

\begin{figure}[h]
    \centering
    {\includegraphics[width=\textwidth]{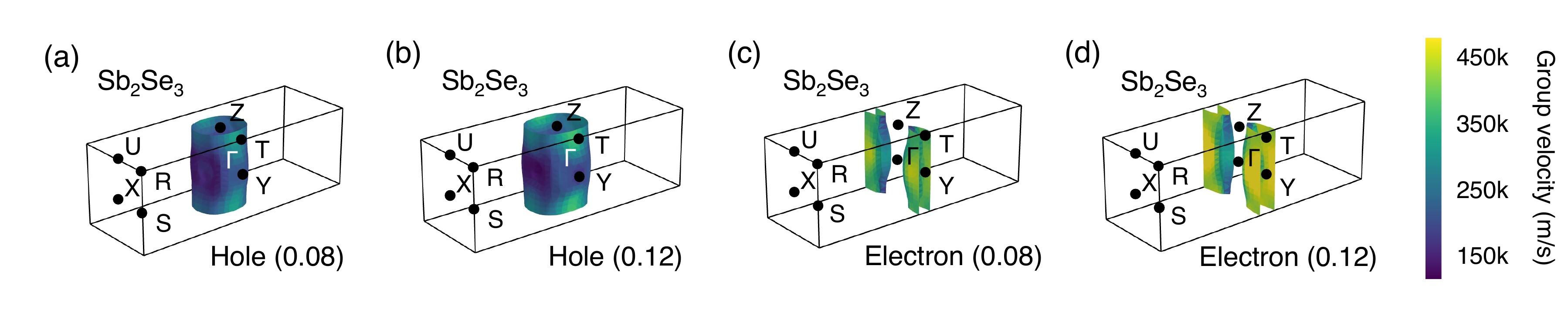}} \\
    \caption{Fermi surfaces of \SbSe. (a) and (b) are hole Fermi surfaces (0.08 and 0.12 eV below the VBM, respectively), while (c) and (d) are electron Fermi surfaces (0.08 and 0.12 eV above the CBM, respectively). The different colors represent the magnitude of group velocity (m/s)}
    \label{fig_Fermi_Se}
\end{figure}

Electron Fermi surfaces (0.08 and 0.12 eV above the CBM) and hole Fermi surfaces (0.08 and 0.12 eV below the VBM) of \SbSsp and \SbSesp are shown in Fig. \ref{fig_Fermi_S} and \ref{fig_Fermi_Se}, respectively. The shapes are qualitatively consistent with the results obtained by 0.10 eV above (below) the CBM (VBM), indicating our conclusions are robust.

\section{Convergence tests} 

\subsection{Total energies}

Convergence tests of total energies with respect to \textit{k}-point density were performed using the optB86b-vdW functional with a plane-wave energy cutoff of 400 eV. As shown in Table \ref{tab_energies}, total energies of \SbSsp and \SbSesp converge to within 1 meV/atom at a \textit{k}-point mesh of 7$\times$2$\times$2.

\begin{table}[h]
\begin{tabular}{ccc}
\hline
System & \textit{k}-point grid & Total energy \\ \hline
\multirow{3}{*}{\ce{Sb2S3}} & 7$\times$2$\times$2 & -56.164 \\
 & 10$\times$3$\times$3 & -56.167 \\
 & 14$\times$4$\times$4 & -56.168 \\ [0.5ex]
\multirow{3}{*}{\ce{Sb2Se3}} & 7$\times$2$\times$2 & -47.126 \\
 & 10$\times$3$\times$3 & -47.131 \\
 & 14$\times$4$\times$4 & -47.132 \\ \hline
\end{tabular}
\caption{Total energies (eV) per unit cell of \SbSsp and \SbSesp with respect to \textit{k}-point density}
\label{tab_energies}
\end{table}

\subsection{Effective masses}

The convergence of effective masses were performed using the HSE06 functional and D3 dipersion correction as shown in Table \ref{tab_eff_test}. The results are sensitive to \textit{k}-point mesh especially along the \textit{z} direction as a result of the non-parabolicity. A \textit{k}-point mesh of 19$\times$10$\times$10 was used for calculations of effective masses in \SbSsp and \SbSe.

\begin{table}[h]
\centering
\resizebox{\textwidth}{!}{%
\begin{tabular}{cccccccccc}
\hline
\multirow{2}{*}{System} & \multirow{2}{*}{\textit{k}-point grid} & \multicolumn{2}{c}{x} & \multicolumn{2}{c}{y} & \multicolumn{2}{c}{z} & \multicolumn{2}{c}{$\overline{m}^{*}$} \\ \cline{3-10}
& & \textit{m}$_{e}^{*}$/\textit{m}$_{0}$ & \textit{m}$_{h}^{*}$/\textit{m}$_{0}$ & \textit{m}$_{e}^{*}$/\textit{m}$_{0}$ & \textit{m}$_{h}^{*}$/\textit{m}$_{0}$ & \textit{m}$_{e}^{*}$/\textit{m}$_{0}$ & \textit{m}$_{h}^{*}$/\textit{m}$_{0}$ & \textit{m}$_{e}^{*}$/\textit{m}$_{0}$ & \textit{m}$_{h}^{*}$/\textit{m}$_{0}$ \\ \hline
\multirow{6}{*}{\ce{Sb2S3}} & 14$\times$4$\times$4 & 0.12 & 0.39 & 1.18 & 0.67 & 7 & 1.17 & 0.32 & 0.61 \\
 & 15$\times$6$\times$6 & 0.16 & 0.42 & 1.10 & 0.67 & 5 & 0.62 & 0.41 & 0.55 \\
 & 16$\times$7$\times$7 & 0.17 & 0.42 & 1.25 & 0.66 & 6 & 0.98 & 0.44 & 0.61 \\
 & 17$\times$8$\times$8 & 0.18 & 0.45 & 1.04 & 0.66 & 5 & 0.97 & 0.45 & 0.63 \\
 & 18$\times$9$\times$9 & 0.16 & 0.46 & 1.09 & 0.65 & 6 & 0.97 & 0.41 & 0.63 \\
 & 19$\times$10$\times$10 & 0.16 & 0.47 & 0.92 & 0.65 & 5 & 0.97 & 0.40 & 0.64 \\ [1ex]
\multirow{6}{*}{\ce{Sb2Se3}} &14$\times$4$\times$4 & 0.11 & 0.89 & 0.86 & 0.73 & 7 & 3 & 0.29 & 1.06 \\
&15$\times$6$\times$6 & 0.12 & 0.87 & 0.79 & 0.57 & 5 & 3 & 0.31 & 0.93 \\
&16$\times$7$\times$7 & 0.13 & 0.90 & 0.98 & 0.56 & 7 & 5 & 0.34 & 0.97 \\
&17$\times$8$\times$8 & 0.14 & 0.94 & 0.84 & 0.52 & 6 & 4 & 0.35 & 0.93 \\
&18$\times$9$\times$9 & 0.14 & 0.89 & 1.00 & 0.53 & 8 & 5 & 0.36 & 0.93 \\
&19$\times$10$\times$10 & 0.14 & 0.85 & 0.81 & 0.55 & 7 & 3 & 0.35 & 0.90 \\ \hline
\end{tabular}%
}
\caption{Effective masses of \SbXsp with respect to \textit{k}-point meshes. Values larger than 2 are rounded to the nearest whole numbers. The harmonic mean is represented by $\overline{m}^{*}$.}
\label{tab_eff_test}
\end{table}

\newpage
\subsection{Dielectric constants}

The convergence of high-frequency dielectric constants (\textit{\textepsilon}$_{\infty}$) is usually sensitive to \textit{k}-point density and number of bands, while the convergence of ionic dielectric constants (\textit{\textepsilon}$_\textrm{ionic}$) is always related to \textit{k}-point density and plane-wave energy cutoff \cite{hybertsen1987ab,olevano1999exchange}.
Convergence tests of dielectric constants (shown in Table \ref{tab_convergence_h} and \ref{tab_convergence_i}) were performed using the optB86b-vdW functional. High-frequency dielectric constants converge to $\sim$0.1 when the number of bands is 128 and \textit{k}-point meshes are 12$\times$4$\times$4 (15$\times$6$\times$6) for \SbSsp(\SbSe). The results of high-frequency dielectric constants in the main text were obtained by the HSE06 functional and D3 correction using the converged parameters shown above. Ionic dielectric constants converge to within 0.1 when the plane-wave energy cutoff is 700 (600) and \textit{k}-point mesh is 15$\times$6$\times$6 (16$\times$7$\times$7) for \SbSsp (\SbSe), which are the settings used for our results.

Table \ref{tab_frequency} shows vibrational frequencies extracted from DFPT phonon calculations at the $\varGamma$ point which are used for calculating ionic contribution to dielectric constants. There are 3 acoustic phonon modes and 57 optical phonon modes with no imaginary phonon modes, indicating our calculations are solid.

For optical absorption coefficients calculations, \textit{k}-point mesh was set to 15$\times$6$\times$6 and the number of bands was set to 160, as the optical absorption coefficients were calculated from high-frequency dielectric constants and they 
\begin{table}[p]
\centering
\resizebox{\textwidth}{!}{%
\begin{tabular}{@{\extracolsep{0.1cm}}ccccccccc}
\hline
\multirow{2}{*}{System} &\multirow{2}{*}{Number of bands} &\multirow{2}{*}{\textit{k}-point grid} & \multicolumn{3}{c}{\textit{\textepsilon}$_{\infty}$} & \multicolumn{3}{c}{Difference} \\  \cline{4-9}
& & & x & y & z & x & y & z \\ \cline{1-9} 
\multirow{9}{*}{\ce{Sb2S3}} &\multirow{4}{*}{128} &6$\times$2$\times$2 & 16.79	& 17.33	& 13.07 &  &  & \\
& &10$\times$3$\times$3 & 17.66	& 17.29	& 12.89	& -0.87	& 0.05	& 0.18 \\
& &12$\times$4$\times$4 & 17.57	& 17.24	& 12.87	& 0.08	& 0.05	& 0.02  \\
& &15$\times$6$\times$6 & 17.60	& 17.27	& 12.89	& -0.03	& -0.03	& -0.02 \\ \cline{2-9}

&68 &\multirow{5}{*}{12$\times$4$\times$4} & 15.18 & 14.87 & 9.93 &  &  &  \\ 
&88 & & 17.08 & 16.70 & 12.28 & -1.90 & -1.83 & -2.36 \\ 
&108 & & 17.48 & 17.13 & 12.76 & -0.40 & -0.43 & -0.48 \\ 
&128 & & 17.57 & 17.24 & 12.87 & -0.09 & -0.11 & -0.11 \\ 
&148  && 17.61 & 17.27 & 12.91 & -0.04 & -0.04 & -0.03\\ \cline{1-9}

\multirow{9}{*}{\ce{Sb2Se3}}&\multirow{4}{*}{128} &6$\times$2$\times$2 &22.29	&26.17	&18.00 & & & \\
& &10$\times$3$\times$3 &24.30	&25.86	&17.58	&-2.00	&0.31	&0.31\\
& &12$\times$4$\times$4 &24.08	&25.94	&17.57	&0.22	&-0.08	&-0.08\\
& &15$\times$6$\times$6 &24.12	&25.99	&17.60	&-0.05	&-0.05	&-0.05\\ \cline{2-9}

&68 &\multirow{5}{*}{15$\times$6$\times$6} & 21.70	& 23.57	& 14.40 &  &  &  \\
&88 & & 23.64	& 25.47	& 17.04	& -1.94	& -1.89	& -2.65 \\
&108 & & 24.04	& 25.89	& 17.50	& -0.39	& -0.42	& -0.46 \\ 
&128 & & 24.13	& 25.99	& 17.60	& -0.09	& -0.11	& -0.10 \\ 
&148 & & 24.17	& 26.02	& 17.64	& 0.04	& -0.03	& -0.04\\ 
\hline
\end{tabular}%
}
\caption{High-frequency dielectric constants (\textit{\textepsilon}$_{\infty}$) of \SbXsp with respect to number of bands and \textit{k}-point meshes}
\label{tab_convergence_h}
\end{table}
\begin{table}[p]
\centering
\resizebox{\textwidth}{!}{%
\begin{tabular}{@{\extracolsep{0.1cm}}ccccccccc}
\hline
\multirow{2}{*}{System} &\multirow{2}{*}{Energy cutoff} &\multirow{2}{*}{\textit{k}-point grid} & \multicolumn{3}{c}{\textit{\textepsilon}$_{\infty}$} & \multicolumn{3}{c}{Difference} \\  \cline{4-9}
& & & x & y & z & x & y & z \\ \cline{1-9} 
\multirow{6}{*}{\ce{Sb2S3}} &\multirow{3}{*}{600} &12$\times$4$\times$4 & 87.42	&83.33	&4.89 &  &  & \\
& &15$\times$6$\times$6 & 87.36	& 83.26	& 4.89	& 0.06	& 0.07	& 0.00\\
& &16$\times$7$\times$7 & 87.34	& 83.22	& 4.89	& 0.02	& 0.04	& 0.00  \\ \cline{2-9}

&400 &\multirow{4}{*}{15$\times$6$\times$6} & 87.41	&82.86	&4.88 &  &  &  \\ 
&500 & & 87.42	&83.10	&4.88	&-0.01	&-0.24	&0.00 \\ 
&600 & & 87.36	&83.26	&4.89	&0.06	&-0.16	&-0.01 \\
&700 & & 87.39	&83.24	&4.89	&-0.03	&0.02	&0.00 \\  \cline{1-9}

\multirow{6}{*}{\ce{Sb2Se3}} &\multirow{3}{*}{600} &12$\times$4$\times$4 & 70.68	&113.64	&4.47 &  &  & \\
& &15$\times$6$\times$6 & 70.53	&113.32	&4.47	&0.15	&0.32	&0.00 \\
& &16$\times$7$\times$7 & 70.53	& 113.26	& 4.47	& 0.00	& 0.06	& 0.00 \\ \cline{2-9}

&400 &\multirow{3}{*}{15$\times$6$\times$6} & 70.54	&112.98	&4.47 &  &  &  \\ 
&500 & & 70.53	&113.25	&4.47	&0.01	&-0.28	&0.00 \\ 
&600 & & 70.53	&113.32	&4.47	&0.00	&-0.07	&0.00 \\ 
\hline
\end{tabular}%
}
\caption{Ionic dielectric constants (\textit{\textepsilon}$_\textrm{ionic}$) of \SbXsp with respect to plane-wave energy cutoffs and \textit{k}-point meshes}
\label{tab_convergence_i}
\end{table}
will show the same convergence behaviour. The optical absorption coefficients were calculated using tetrahedron method with Bl\"{o}chl corrections.

\begin{table}[ht]
\small
\centering
\begin{tabular}{@{\extracolsep{0.2cm}}cccccc}
\hline
\multirow{2}{*}{Normal mode} & \multicolumn{2}{c}{Vibrational frequency}& \multirow{2}{*}{Normal mode}& \multicolumn{2}{c}{Vibrational frequency}\\\cline{2-3} \cline{5-6}
&{\ce{Sb2S3}} & {\ce{Sb2Se3}} &  & {\ce{Sb2S3}} & {\ce{Sb2Se3}}\\ \hline
1 & 6.38 & 9.67 & 31 & 3.19 & 4.84 \\
2 & 6.19 & 9.19 & 32 & 3.16 & 4.84 \\
3 & 5.95 & 8.88 & 33 & 2.94 & 3.90 \\
4 & 5.93 & 8.82 & 34 & 2.86 & 3.83 \\
5 & 5.82 & 8.80 & 35 & 2.79 & 3.78 \\
6 & 5.75 & 8.74 & 36 & 2.72 & 3.71 \\
7 & 5.62 & 8.52 & 37 & 2.68 & 3.59 \\
8 & 5.33 & 8.34 & 38 & 2.58 & 3.43 \\
9 & 5.28 & 8.32 & 39 & 2.44 & 3.33 \\
10 & 5.21 & 7.86 & 40 & 2.27 & 2.98 \\
11 & 5.15 & 7.60 & 41 & 2.14 & 2.81 \\
12 & 5.13 & 7.50 & 42 & 1.90 & 2.52 \\
13 & 4.99 & 7.47 & 43 & 1.88 & 2.35 \\
14 & 4.90 & 7.44 & 44 & 1.79 & 2.20 \\
15 & 4.61 & 7.28 & 45 & 1.70 & 2.19 \\
16 & 4.44 & 7.19 & 46 & 1.69 & 2.00 \\
17 & 4.25 & 6.79 & 47 & 1.65 & 1.90 \\
18 & 4.22 & 6.73 & 48 & 1.61 & 1.89 \\
19 & 4.21 & 6.49 & 49 & 1.58 & 1.70 \\
20 & 4.14 & 6.42 & 50 & 1.57 & 1.66 \\
21 & 3.87 & 6.11 & 51 & 1.38 & 1.54 \\
22 & 3.86 & 6.03 & 52 & 1.34 & 1.47 \\
23 & 3.81 & 6.03 & 53 & 1.31 & 1.39 \\
24 & 3.81 & 5.98 & 54 & 1.23 & 1.35 \\
25 & 3.71 & 5.97 & 55 & 1.18 & 1.27 \\
26 & 3.69 & 5.82 & 56 & 0.86 & 1.03 \\
27 & 3.52 & 5.69 & 57 & 0.52 & 0.75 \\
28 & 3.49 & 5.62 & 58 & 0 & 0 \\
29 & 3.34 & 5.61 & 59 & 0 & 0 \\
30 & 3.31 & 5.39 & 60 & 0 & 0 \\ \hline
\end{tabular}
\caption{Vibrational frequencies (THz) of \SbXsp at the $\varGamma$ point}
\label{tab_frequency}
\end{table}

~\\
\newpage
\bibliography{references} 
\bibliographystyle{rsc}